\documentclass[useAMS,usenatbib,usedcolumn,usegraphicx]{mn2e}

 \usepackage[english]{babel}
 \usepackage[T1]{fontenc}
 \usepackage{times}
 \usepackage{amssymb}
 \usepackage{amsmath}				%
 \usepackage{multirow}
 \usepackage[
	pdftex,
	pdfpagemode={UseOutlines},
	bookmarks,
	bookmarksopen,
	colorlinks,
	linkcolor={blue},
	citecolor={green},
	urlcolor={red}
]{hyperref}
\graphicspath{{ART/}}
 \DeclareGraphicsExtensions{.pdf,.png,.jpg}
\def\smartspace#1{{\protect\aftergroup\smartspaceit#1}}
\def\smartspaceit{\futurelet\spta\sptest}
\def\sptest{\ifcat\noexpand\spta,\else\ \fi}
\newcommand{\nc}{\newcommand}

\newcommand{\esm}{\ensuremath}
\newcommand{\mcl}{\multicolumn}
\newcommand{\sspc}{\smartspace}
\nc{\etal}{\sspc{et~al.}}
\nc{\eg}{\sspc{e.g.}}
\nc{\etc}{\sspc{etc.}}
\nc{\cf}{\sspc{cf.}}
\nc{\ie}{\sspc{i.e.}}
\nc{\cfig}[1]{\centerline{\psfig{#1}}}
\nc{\Ncol}[1]{\mcl{#1}{c}{~}}
\nc{\mcn}[1]{\mcl{#1}{l}{~}}
\nc{\mcN}[2]{\mcl{#1}{c}{#2}}
\nc{\mcc}[1]{\mcl{1}{c}{#1}}
 \nc{\AAA}{\sspc{\esm{\lambda\lambda}}}
 \nc{\amm}{\sspc{\,\AA\,mm$^{-1}$}}
 \nc{\km}{\sspc{\,\esm{\text{km}}}}
 \nc{\msec}{\sspc{\,\esm{\text{ms}}}}
 \nc{\kms}{\sspc{\,\esm{\text{km\,s}^{-1}}}}
 \nc{\masyr}{\sspc{\,\esm{\text{mas\,yr}^{-1}}}}
 \nc{\msun}{\sspc{\,\esm{\text{M}_{\odot}}}}
 \nc{\rsun}{\sspc{\,\esm{\text{R}_{\odot}}}}
 \nc{\lsun}{\sspc{\,\esm{\text{L}_{\odot}}}}
 \nc{\yr}{\sspc{\,\esm{\text{yr}}}}
 \nc{\kyr}{\sspc{\,\esm{\text{kyr}}}}
 \nc{\pc}{\sspc{\,\esm{\text{pc}}}}
 \nc{\kpc}{\sspc{\,\esm{\text{kpc}}}}
 \nc{\halpha}{\sspc{\,\mbox{H$\alpha$}}}
 \nc{\sigmad}{\sspc{$\Sigma-$\text{D}}}
 \nc{\Av}{\sspc{\esm{\text{A}_\text{V}}}}
 \nc{\Mv}{\sspc{\esm{\text{M}_\text{V}}}}
 \nc{\ion}[2]{\sspc{#1\,{\scshape#2}}}
 \nc{\dgr}{\esm{^\circ}}
 \nc{\asec}{\sspc{\,\raisebox{0.25ex}{\scshape"}}}
 \nc{\Mag}[2]{#1\fm#2}
 \nc{\dex}[1]{\esm{10^{#1}}}
 \nc{\tdex}[1]{\esm{\times10^{#1}}}
 \nc{\pmpm}[2]{\sspc{\esm{^{#1}_{#2}}}}
 \nc{\todo}[1]{\sspc{(\textbf{TODO:} #1)}}
 \nc{\obj}[1]{\sspc{#1}}
 \nc{\rotsed}{\sspc{ROTSE--IIId}}
 \nc{\rotse}{\sspc{ROTSE--III}}
 \nc{\tubitak}{\sspc{T\"UB\.ITAK}}

\title
[%
  Runaway in S147%
]{%
  Discovery of an OB Runaway Star Inside SNR S147%
}
\author
[%
  Din\c{c}el et al%
]{%
  B.\     Din\c{c}el$^{1}$\thanks{Correspondence address: baha.dincel@uni-jena.de},
  R.\     Neuh\"auser$^{1}$,
  S.\,K.\ Yerli$^{2}$,
  A.\     Ankay$^{3}$,
  N.\     Tetzlaff$^{1}$,
  G.\     Torres$^{4}$,
  M.\     Mugrauer$^{1}$ \\
  $^{1}$ Astrophysikalisches Institut und Universit\"{a}ts-Sternwarte Jena, 07745 Jena, Germany\\
  $^{2}$ Orta Do\u{g}u Teknik \"Universitesi, Department of Physics, 06531 Ankara, Turkey\\
  $^{3}$ Bo\u{g}azi\c{c}i University, Department of Physics, 34342 \.{I}stanbul, Turkey\\
  $^{4}$ Harvard-Smithsonian Center for Astrophysics, 60 Garden St., Mail Stop 20, Cambridge MA 02138, USA%
}
\begin{document}
\date{%
	Accepted 2015 January 17.
	Received 2014 December 12;
	in original form 2014 May 25
}
\volume{000}
\pagerange{\pageref{firstpage}--\pageref{lastpage}}
\pubyear{2014}

\maketitle
\label{firstpage}
\begin{abstract}
We present first results of a long term study: Searching for OB--type runaway stars inside supernova remnants (SNRs).
We identified spectral types and measured radial velocities (RV) by optical spectroscopic observations and
we found an early type runaway star inside SNR S147.
HD 37424 is a B0.5V type star with a peculiar velocity of 74$\pm$8 \kms.
Tracing back the past trajectories via Monte Carlo simulations,
we found that HD 37424 was located at the same position as the central compact object, PSR J0538+2817, $30\!\pm\!4$\,\kyr ago.
This position is only $\sim$4 arcmin away from the geometrical center of the SNR.
So, we suggest that HD 37424 was the pre--supernova binary companion to the progenitor of the pulsar and the SNR.
We found a distance of 1333$\pmpm{+103}{-112}$ \pc to the SNR.
The zero age main sequence progenitor mass should be greater than 13 \msun.
The age is $30\pm4$ \kyr and the total visual absorption towards the center is 1.28$\pm$0.06 mag.
For different progenitor masses, we calculated the pre--supernova binary parameters.
The Roche Lobe radii suggest that it was an interacting binary in the late stages of the progenitor.
\end{abstract}


\section{Introduction}

High space velocities of OB runaway stars are explained by two independent mechanisms:
dynamical ejection due to gravitational interactions of massive stars in cluster cores
	\citep{1967BOTT....4...86P}
and binary disruption as a result of a supernova explosion of the initially more massive component
	\citep{1961BAN....15..265B}.
Both scenarios are viable, but whether one of the mechanisms is dominant is still uncertain.
According to the virial theorem, through a symmetric supernova explosion in a binary system, if more than half of the total mass of the system is released,
then the new born neutron star (or black hole) and the non--degenerate component are no more gravitationally bound
	\citep{1961BAN....15..265B}.
However, the energy stored in the orbit, in most cases, is not sufficient to produce the neutron star kick velocities that are typically in the range of 300--500 \kms
	\citep{%
		1994Natur.369..127L,
		1997ARep...41..257A,
		2005MNRAS.360..974H,
		1997MNRAS.291..569H}.
The asymmetry in supernova explosions is responsible for such high velocities
        \citep{2013A&A...552A.126W}.
Therefore, there are no pulsar companions to many of the OB runaway stars
	\citep{1996ApJ...461..357S}.
In some cases, the compact object does not receive a significant kick and/or the majority of the total mass is stored on the secondary through conservative mass transfer, hence, the compact object remains bound to the companion star
\citep{1993SSRv...66..309V}.
The runaway high mass X--ray binaries like 4U1700-37 \citep{2001A&A...370..170A} and Vela X--1 \citep{1997ApJ...475L..37K} are such examples.

Yet, the low rate of X--ray binaries and the high rate of isolated neutron stars
by taking into consideration the selection effects, the binary disruption is likely to occur in most cases
	\citep{2005Ap.....48..330G} (hereafter, G05).
The kinematics of the binary disruption due to an asymmetric supernova explosion is widely discussed in
	\citet{1998A&A...330.1047T}.
However, a sample of observationally confirmed OB runaway--NS couples is needed for a better understanding of the problem.

The importance of searching for OB runaways inside SNRs was first mentioned in
	\citet{1980JApA....1...67V}.
However, the kinematical study of known OB stars inside SNRs was concluded with a lack of OB runaways due to the poor sample of SNRs and OB stars.
Still, there is no known O or B--type runaway star that can be directly linked to an SNR given in the literature.

In G05, the outcome of exploring runaway pairs from binary supernova disruption is broadly discussed.
Firstly, identifying the explosion centers more precisely will be useful for determining the velocities of young NSs
of which proper motion measurements have high uncertainties.
Thus, the kick that is gained by the NS due to the asymmetry of the SN can be determined more precisely.
Secondly, the distance to the remnant can be measured more accurately by studying the runaway star,
as it cannot move far away from the explosion center in the observational lifetime of the SNR.
Finally, possible effects of a close binary system on the asymmetry of the SNe can also be examined.
Additionally, it would be a direct evidence of the binary supernova scenario (BSS).

The observational efforts are concentrated on runaway--NS coupling and abundance investigations of runaway stars.
Some examples of the runaway--NS pairs (components are separated) are PSR\,B1929+10 -- $\zeta$ Oph
	\citep{%
               2001A&A...365...49H,
	       2008AstL...34..686B,
               2010MNRAS.402.2369T},
PSR\,J0630-2834 -- HIP\,47155
        \citep{2013MNRAS.435..879T}
and PSR\,J0826+2637 -- HIP13962
        \citep{2014MNRAS.438.3587T}.
Based on their motions in space, the pulsars and the corresponding runaway stars were traced back in time by using 3--D Monte Carlo simulations.
They were found at the same position and the same time inside a young open cluster.
The PSR\,J0630-2834 -- HIP\,47155 pair is thought to be ejected from very old SNR Antlia.

There are considerable uncertainties in these cases as the supernova events took place more than \dex{5} yr ago.
The separation between the objects is very large and the SNR has faded away long ago and/or the components are outside of the SNR (the Antlia case).
Other important observational evidence is the enhancement of $\alpha$--process elements in the hyper--velocity star HD\,271791.
The star is proposed to be ejected from a massive close binary system due to an SN
that enriches its photosphere in elements which can be synthesized in large amounts during the evolution of the progenitor
	\citep{2008ApJ...684L.103P}.

The method followed in our work is a direct study of possible runaway stars inside SNRs as described in G05.

Briefly, assuming that the massive binary mass ratio is greater than 1:4,
OB--type star candidates are determined by a careful study of their BVJHK magnitudes obtained from the UCAC4 catalog
\citep{2012yCat.1322....0Z}.
The angular separation of the sources from the geometrical centers of the remnants are within the limits of the maximum angular distance
that a runaway star can have in a lifetime of an SNR.
The distance moduli are calculated for all sources within this region,
and the sources having extinctions and radial distances consistent with those of the SNRs are considered as "candidates".
Measuring the radial velocities and identifying the spectral types of these objects via spectroscopy
reveal their runaway nature, their youth and the exact spatial relations with the SNRs, in other words, the genetic connections.

Although the runaway stars arising from BSS may also be late type stars,
it is time consuming to check the possible runaway nature of all the stars inside each SNR.
Also, the late type stars are too faint to observe at large distances.
A star having the same age as a supernova progenitor must be young.
As OB type stars evolve faster, they automatically satisfy this condition.
Furthermore, high mass stars are rare objects.
An OB runaway star discovered inside an SNR can be explained by the BSS.
Considering the very short observable lifetimes of SNRs and relatively short later evolution stages of stars,
most probably, main sequence stars are expected as OB runaways connected to SNRs.

The space velocity of an OB runaway star is thought to be larger than 30$-$40 \kms
	\citep{%
		1965MNRAS.130..245F,
		1996ApJ...461..357S}.
A more precise value is proposed in
	\citet{2011MNRAS.410..190T}; 28 \kms in 3--D and 20 \kms in 2--D.
To summarize, a main--sequence OB type star of
which at least one component of the velocity vector is greater than {20} \kms is searched in selected SNRs.

The first criterion of the candidate selection is the restriction of the angular position.
Most of the OB runaways have peculiar velocities lower than 80 \kms
	\citep{%
		1986ApJS...61..419G,
		1996AJ....111.1220P,
                2011MNRAS.410..190T}.
As the shock wave velocity of the SNRs are decelerated from roughly 10000 \kms to several hundred \kms,
it is expected that the runaway star cannot exceed one tenth of the angular diameter ($\theta$) of the related SNR.
This value is somewhat relaxed to $\theta$/6 considering the uncertainties in the geometrical centers of the SNRs.
The stars in this region are expected to be consistent with the respective SNR in terms of distance and reddening.

For this comparison, the adopted distances of SNRs given in
	\citet{%
		2003SerAJ.167...93G,
		2004SerAJ.168...55G,
		2004SerAJ.169...65G},
and \Av values from
	\citet{1980BICDS..19...61N}
were used.
48 SNRs within 5 \kpc from the Sun were selected for investigation.

\begin{figure*}
\includegraphics[width=10cm,height=9cm]{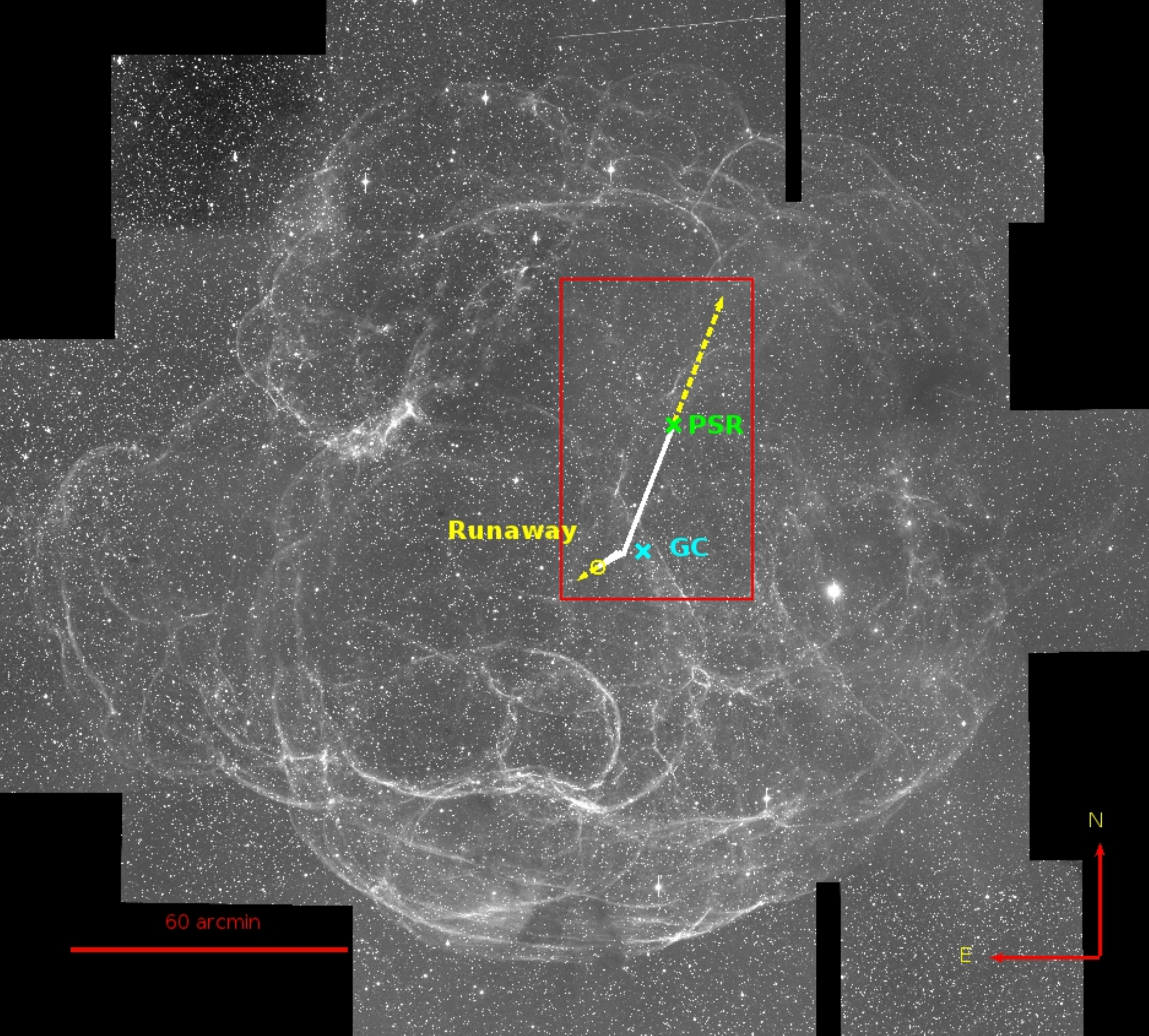}
\includegraphics[width=6cm,height=9cm]{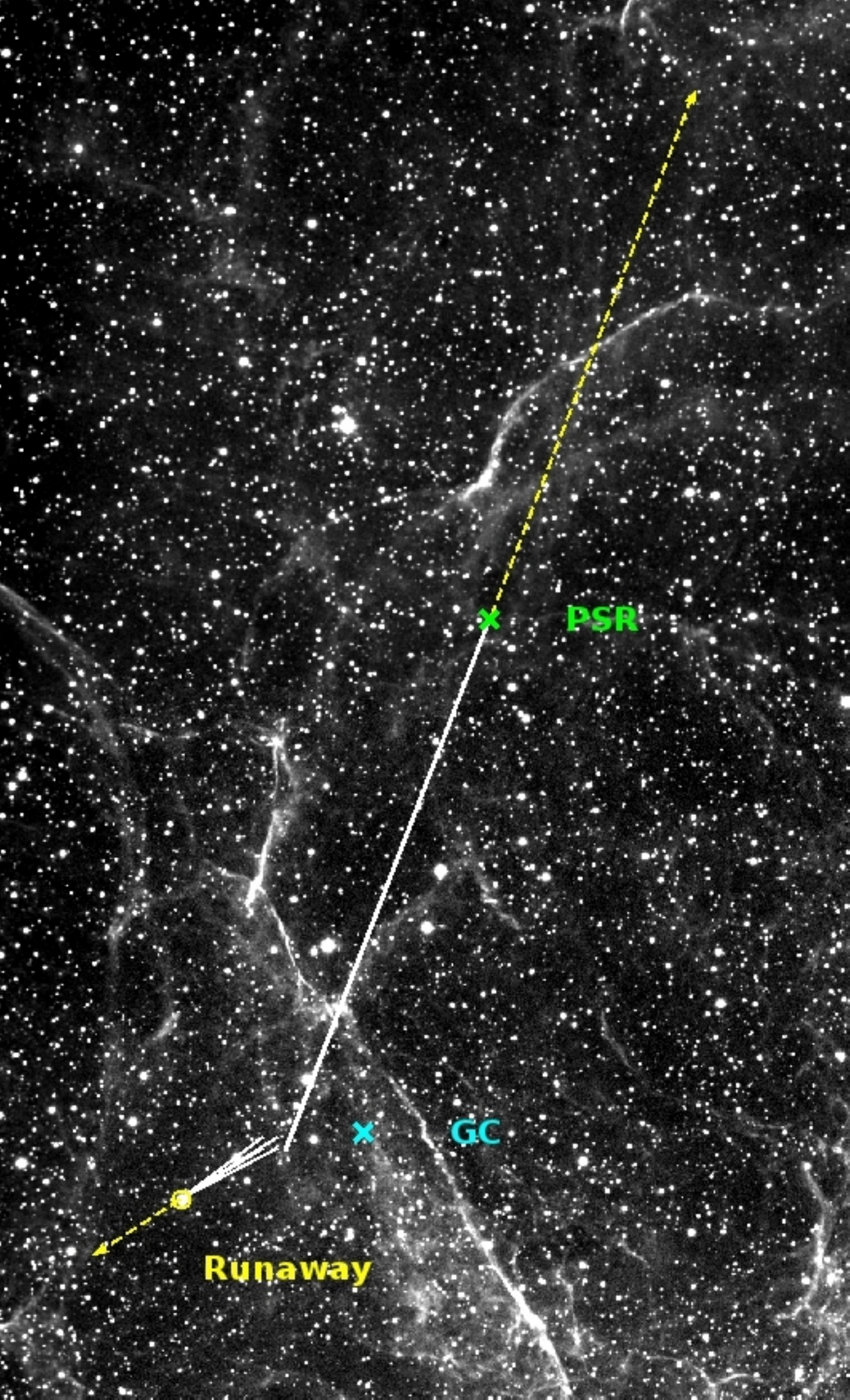}
\caption{
The $4.8\,^{\circ}\times4.1\,^{\circ}$
\halpha image of SNR S147 taken at the University Observatory Jena.
The green cross represents the pulsar, the cyan cross shows the GC of the SNR  and the yellow circle is HD\,37424.
The yellow vectors show the proper motion of the objects.
The red box is the zoom--in area shown in the right panel.
The white arrows are the tracing back cones of the proper motion for 29300 yr.
The angle between the pm vectors is $139^{\circ}-148^{\circ}$
In 2--D calculations, both objects come towards each other as close as 0.126 \pc which means that they have a common origin; binary supernova disruption.}
\label{f:mosaic}
\end{figure*}

In this paper, the result of the runaway search in SNR G180.0-1.7 (S147) is given.
The kinematic relation between the runaway star HD\,37424 and the pulsar PSR\,J0538+2817 is shown,
the possible host OB association is discussed,
the SNR parameters are constrained and
the pre--supernova binary is constructed.
\section{S147, PSR\,J0538+2817 and HD\,37424}

S147 is a shell type SNR located in the Galactic anti--center direction.
It is 180 arcmin in diameter with a geometrical center (GC) at
$\alpha=05\mathrm{h} 39\mathrm{m} 00\mathrm{s}$, $\delta=+27\,^{\circ} 50\mathrm{'} 00\mathrm{''}$
\citep{2009yCat.7253....0G}.
It was first mentioned as an SNR candidate in
\citet{1958RvMP...30.1048M}.
The compact object related to the SNR is radio pulsar PSR\,J0538+2817
\citep{1996ApJ...468L..55A}.
In optical bands, the shell structure is well defined and dominated by filamentary emission in \halpha.
The emission is brighter in the north and south edges and mainly concentrated in the southern parts.
Despite of its old age, it conserves the spherical symmetry except for the blowout regions in east and west
(Figure \ref{f:mosaic}).
The total absorption in the V band is \Av=$0.7\pm0.2$ magnitude
\citep{1985ApJ...292...29F}.
Radio observations reveal that the spectral index is unusually varying.
The shell structure observed at radio wavelengths coincides with that in optical bands and is well defined
\citep{1986A&A...163..185F}.
But, no X--ray emission is observed from the remnant
\citep{1990A&A...227..183S}.
Suggested distances vary from 0.6 to 1.9 \kpc in various publications (Table \ref{t:snr_distance}).
Four of the measurements are based on \sigmad relation which is useful to estimate the distance by using the surface brightness of the SNR.
In \citet{1980PASJ...32....1S} and \citet{1980A&A....92..225K}, the model of
\citet{1979AuJPh..32...83M}
is used, while in \citet{1976MNRAS.174..267C} and \citet{2003A&AT...22..273G}, estimations are based on their own models.
The radius lower limit calculated through SNR dynamics based on the model of
\citet{1974ApJ...188..501C}
sets another constraint on the SNR distance.
The distance derived from the pulsar's parallax or dispersion measure is larger than the distance suggested for the background stars of which spectra show high velocity gas related to the SNR.

\begin{table}
\caption{Distance estimates for S147. (R$^{l}$) denotes the radius lower limit of the SNR.
}
\label{t:snr_distance}
\begin{center}
\begin{tabular}{l l l}
Distance (\kpc)
& Method
& Reference \\\hline
1.6$\pm$0.3		 & \sigmad          &  {\citet{1980PASJ...32....1S}} \\
0.8--1.37		 & R$^{l}$, \sigmad &  {\citet{1980A&A....92..225K}} \\
0.6			 & R$^{l}$          &  {\citet{1979ApJ...229..147K}} \\
0.9			 & \sigmad          &  {\citet{1976MNRAS.174..267C}} \\
0.8$\pm$0.1		 & \Av              &  {\citet{1985ApJ...292...29F}} \\
1.06			 & \sigmad          &  {\citet{2003A&AT...22..273G}} \\
0.88<			 & High Vel Gas     &  {\citet{2004A&A...426..555S}} \\
1.2			 & Pulsar DM        &  {\citet{2003ApJ...593L..31K}} \\
1.47$^{+0.42}_{-0.27}$	 & Pulsar Plx       &  {\citet{2007ApJ...654..487N}} \\
1.3$^{+0.22}_{-0.16}$	 & Pulsar Plx       &  {\citet{2009ApJ...698..250C}} \\\hline
\end{tabular}
\end{center}
\end{table}

The age was estimated as more than 100 \kyr
from the Sedov solution by using a shock velocity $\sim$80--100 \kms
\citep{1980PASJ...32....1S,
	   1980A&A....92..225K}.
This velocity was derived from the RV of the filamentary knots in
 \citet{1976SvA....20...19L,
	1979ApJ...229..147K}.
However the pulsar--SNR relation indicates an age of $30\pm4$ \kyr (from the travel time from the GC to the present position)
\citep{2003ApJ...593L..31K,
2007ApJ...654..487N}.
For a distance of 1.3 \kpc, and an angular diameter of 200 arcmin, the radius (R) of the SNR is 38 \pc.
Then, assuming an age of 30 \kyr, the remnant is in its Sedov Phase having a blast wave velocity of 500 \kms.
The estimated explosion energy is between 1$-$3\tdex{51} erg and the corresponding inter--cloud gas density is 0.03$-$0.1 $\mathrm{cm^{-3}}$
\citep{2012ApJ...752..135K}.
The SNR is expanding in a low density medium, probably in the cavity generated by the progenitor.

The central source, PSR\,J0538+2817, is an extensively studied radio and X--ray pulsar located at
$\alpha=05\mathrm{h} 38\mathrm{m} 25.1\mathrm{s}$,
$\delta=+28\,^{\circ} 17\mathrm{'} 09.2\mathrm{''}$,
$\sim$28 arcmin away from the GC towards north.
The 143.16 $\mathrm{ms}$ period and 3.67\tdex{-15} $\mathrm{s\,s^{-1}}$ period derivative
\citep{1996ApJ...468L..55A}
imply a characteristic age of $\sim$620 \kyr which is $\sim$20 times larger than its kinematic age of 30 \kyr.
This discrepancy is explained by either a long initial spin period of P$_{0}=139$ \msec
\citep{2003ApJ...593L..31K}
or strong magnetic field decay
\citep{2004IJMPD..13.1805G}.
The parallax distance was measured as $1.47\pmpm{+0.42}{-0.27}$ \kpc
\citep{2007ApJ...654..487N}
and $1.3\pmpm{+0.22}{-0.16}$ \kpc
\citep{2009ApJ...698..250C}
and the dispersion measure (DM) distance is 1.2 \kpc
\citep{2003ApJ...593L..31K}
by using the NE2001 model of
\citet{2002astro.ph..7156C}
which are consistent with each other.
The most precisely measured proper motion is
$\mu_{\alpha}^{*}=-23.57\pmpm{+0.10}{-0.10}$ \masyr, $\mu_{\delta}=52.87\pmpm{+0.09}{-0.10}$ \masyr
corresponding to a transverse velocity of $357\pmpm{+59}{-43}$ \kms at $1.3\pmpm{+0.22}{-0.16}$ \kpc
\citep{2009ApJ...698..250C}.
Although it presents no clear $\gamma$--ray emission
\citep{2012ApJ...752..135K},
the pulsar is observable in soft X--rays due to thermal emission.
PSR\,J0538+2817 is famous for being one of the few thermal pulsars:
Based on a blackbody model, the surface temperature and the emitting radius are
given as T$=2.12\tdex{6}$ K and R$=1.68\pm0.05$ \km
\citep{2003ApJ...591..380M}.
The hydrogen atmosphere model with B$=1\tdex{12}$ G gives T$=2.12\tdex{6}$ K and R$\simeq$10 \km
\citep{2004MmSAI..75..458Z}.
A faint pulsar wind nebula is observed in X--rays but not in radio
\citep{2000MNRAS.318...58G}.
It is not known whether the observed elongated structure is the torus or the jets of the PWN.
But assuming that it is due to the torus, it provides valuable information on the spin--kick alignment.

HD\,37424 is a sound OB runaway star at 10.3 arcmin away from the GC to the west at
$\alpha=05\mathrm{h} 39\mathrm{m} 44.4\mathrm{s}$, $\delta=+27\,^{\circ} 46\mathrm{'} 51.2\mathrm{''}$.
Photometric and kinematic information on the star is given in Table \ref{t:star}.
While proper motion values were retrieved from the $\textit{UCAC4}$ catalog
\citep{2012yCat.1322....0Z},
the photometric magnitudes are obtained from the $\textit{ASCC-2.5 V3}$ catalog
\citep{2001KFNT...17..409K}.
Its spectral type was previously identified as B0.5/1 IV/V
\citep{1979RA......9..479C}.
As there was no public data nor a visualized spectrum, we established the spectral type again by taking a spectrum.
The star has a very high proper motion compared to other early type and possibly distant stars.
There is no reported variability or binarity in the literature.
Hence, HD\,37424 is a good OB runaway star candidate.

\begin{table}
\caption{BVJHK (in mag) and pm (in \masyr) values of HD\,37424.}
\label{t:star}
\begin{center}
\begin{tabular}{c r@{$\pm$}l}
Parameter
& \multicolumn{2}{c}{Value}	          \\\hline
B			& 9.062&0.017 \\
V			& 8.989&0.019 \\
J			& 8.666&0.017 \\
H			& 8.696&0.026 \\
K			& 8.699&0.019 \\
$\mu_{\alpha}^{*}$	&  10.8&0.8 \\
$\mu_{\delta}$		& -10.2&0.6 \\\hline
\end{tabular}
\end{center}
\end{table}
\section{Observations}

HD\,37424 was observed at the \tubitak National Observatory (TUG)
with the TUG Faint Object Spectrograph and Camera (TFOSC) mounted on the 150\,cm Russian Turkish Telescope (RTT--150).
The observations were carried out on 2013 August 9 and 10.
Two spectra were obtained with a 600\,s exposure time for each.
Grism 14 was used with a 67 micron slit.
The wavelength range in this configuration is 3270--6120 \AA\ and the resolving power (R) is $\sim$1337.
5 He lamp spectra taken in the same night with the object were used for wavelength calibration,
and 5 halogen lamp spectra were obtained in the beginning and at the end of the night for flat--fielding.
Also, 2 more targets had been observed in the same way on 2012 September 13 and 14 for the search for the runaway star.
With the same purpose, 12 objects were observed at the Calar Alto Observatory on 2013 January 29
with Calar Alto Faint Object Spectrograph (CAFOS) on the 2.2 meter telescope.
The covered wavelength range of the grism B--200 is 3200--7000 \AA\ and R$\sim$550.
3 HgCdAr and 3 halogen spectra were taken in the beginning and in the end of the night for wavelength calibration and flat--fielding.
The high resolution observations of HD\,37424 were performed at the Fred L. Whipple Observatory on 2013 September 16 and 21.
The Tillinghast Reflector Echelle Spectrograph (TRES) mounted on the 1.5 meter telescope was used.
Two spectra were taken at R$\sim$44000 in a broad wavelength range; 3800--9100 \AA\ for 300\,s exposure time.
HD\,36665 which is supposed to be a background object to the SNR was observed on 2014 February 11 with the
Fibre Linked Echelle Astronomical Spectrograph (FLECHAS)
\citep{2014AN....335..417M}
at the University Observatory Jena.
The covered wavelength range is 3900--8100 \AA\ and R$\sim$9200.
Three spectra with 600 s exposure time were obtained.
Three tungsten and 3 Th-Ar lamp spectra were taken immediately before this set for flat--fielding and wavelength calibration.

The observations of SNR\,S147 were performed at the University Observatory Jena with the Schmidt-Teleskop-Kamera (STK),
operated in the Schmidt-focus of the 90\,cm telescope of the observatory (see \citet{2010AN....331..449M}).
The observations were carried out with the narrow--band \halpha filter of the STK, which spans a full--width--half--maximum of 5 nm.
In total, 22 fields on the sky were observed during 5 nights, each with a field of view of $52.8'\times52.8'$.
At each position two 600\,s integrations were taken with the STK, i.e. about 7.3 hours of integration time in total.
The individual raw images per field were dark- and flatfield--corrected and finally averaged.
Point sources detected in the overlap regions of the fully processed images were then used to create the mosaic image of SNR\,S147, shown in Figure \ref{f:mosaic},
which covers a total field of view of $4.8^{\circ}\times 4.1^{\circ}$.

FLECHAS, TFOSC and CAFOS data were reduced with IRAF.
All raw frames from TFOSC and CAFOS were over--scan stripped
and no further subtraction was done as long as the over--scanned dark frames yields null count.
Yet, FLECHAS images were subtracted by the dark frames taken in the previous night.
The exposure time of a dark frame is the same as that of the corresponding science or calibration frame.
Hence, the illumination effect was removed.

Then, the flat images were combined while rejecting the frames having minimum and maximum counts.
Next, the object spectra were extracted by assigning an appropriate background region to each.
The arc spectra were extracted as traced by the corresponding aperture of the object spectra.
After wavelength calibration was performed, all spectra were normalized.

\begin{figure}
\includegraphics[width=8cm]{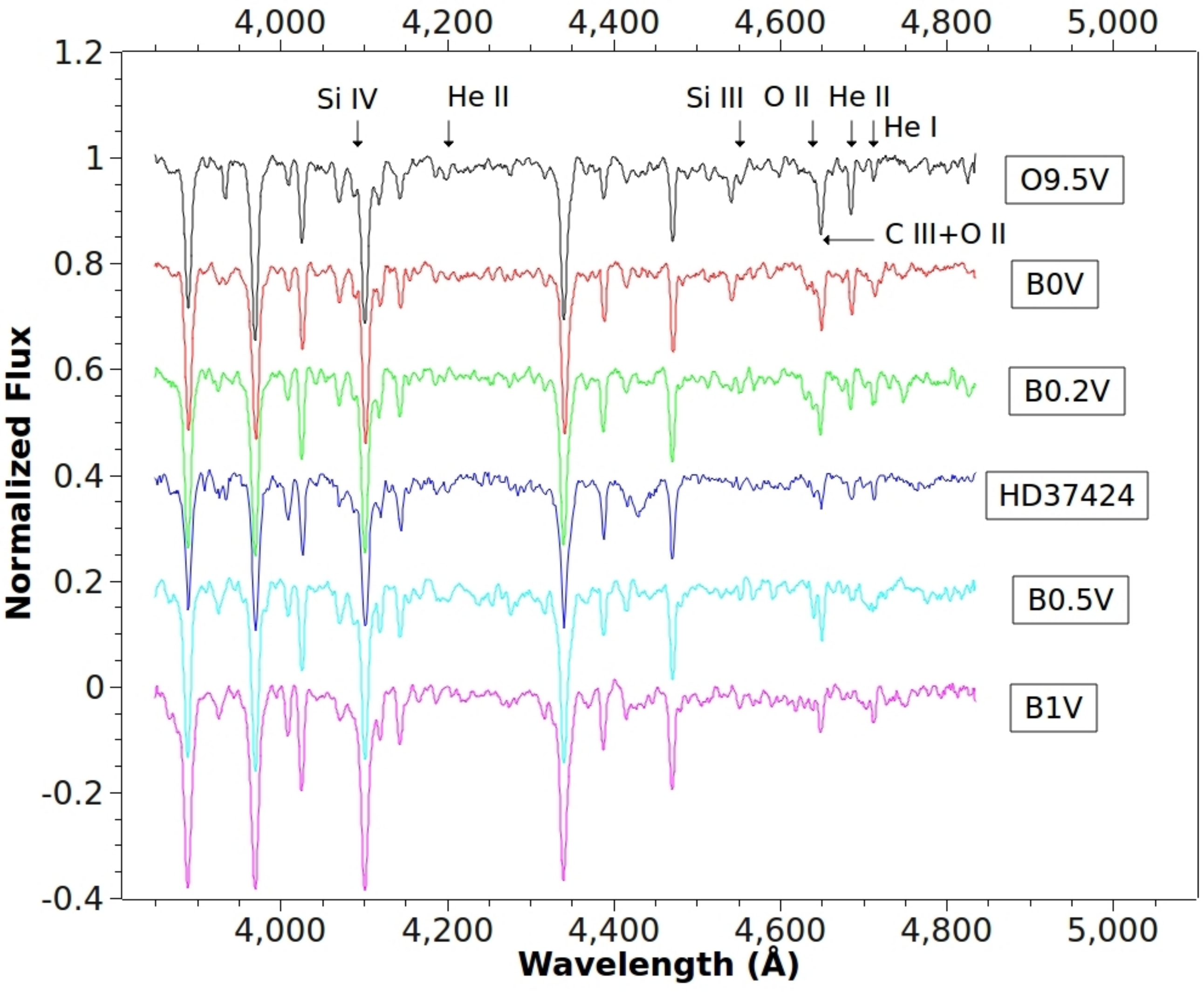}
\caption{TFOSC spectrum of HD\,37424 compared to main sequence B type stars.
Distinctive features are marked. From left to right;
\ion{Si}{iv} $\lambda$4089,
\ion{He}{ii} $\lambda$4200,
\ion{Si}{iii} $\lambda$4552,
\ion{O}{ii} $\lambda$4640,
\ion{C}{iii}+\ion{O}{ii} $\lambda$4650 blend,
\ion{He}{ii} $\lambda$4686,
\ion{He}{i} $\lambda$4713}
\label{f:spt}
\end{figure}

The final spectra were compared with MK standard star spectra retrieved from
\citet{1990PASP..102..379W}.
To maintain equally resolved spectra, the standard spectra were boxcar smoothed by 13.
Figure \ref{f:spt} shows the TFOSC spectrum of HD\,37424 comparing with main sequence B type stars from various subclasses.
HD\,37424 shows
\ion{He}{ii} $\lambda$4686
absorption indicating that the spectral type is earlier than B1.
This feature is not seen in stars with a spectral type later than B0.7.
On the other hand,
\ion{He}{i} lines (\ie \ion{He}{i} $\lambda$4713)
are not as weak as in O type stars compared to
\ion{He}{ii} lines; $\lambda$4200, $\lambda$4686.
The
\ion{Si}{iv} $\lambda$4116
line is totally blended and unmeasurable.
\ion{Si}{iii}  $\lambda$4552
is in low strength compared to
\ion{Si}{iv} $\lambda$4089
like in the B0V type spectrum.
This ratio decreases towards higher temperatures, but
\ion{He}{i} $\lambda$4541
is not stronger than
\ion{Si}{iii} $\lambda$4552
as seen in B0V and O9.5V spectra.
Given the equal strengths of
\ion{O}{ii} $\lambda$4640,
\ion{He}{ii} $\lambda$4686,
\ion{He}{i} $\lambda$4713
and more intense
\ion{C}{iii}+\ion{O}{ii} $\lambda$4650 blends,
HD\,37424 matches best with B0.2V.
Considering the slightly stronger
\ion{He}{i} $\lambda$4713,
B0.5 would be the best approach to the temperature class of the star.
\ion{Si}{iv} $\lambda$4089
strength increases against
\ion{He}{i} $\lambda$4026--4144
as the luminosity class goes higher.
However, neutral helium lines are strong enough to judge that HD\,37424 is a main sequence star;
B0.5V.
The spectral type is between B0.2V--B0.7V, but it is assumed as B0.5$\pm$0.5V in the distance and extinction calculations to avoid underestimation of the errors.

\begin{figure}
\includegraphics[width=8cm]{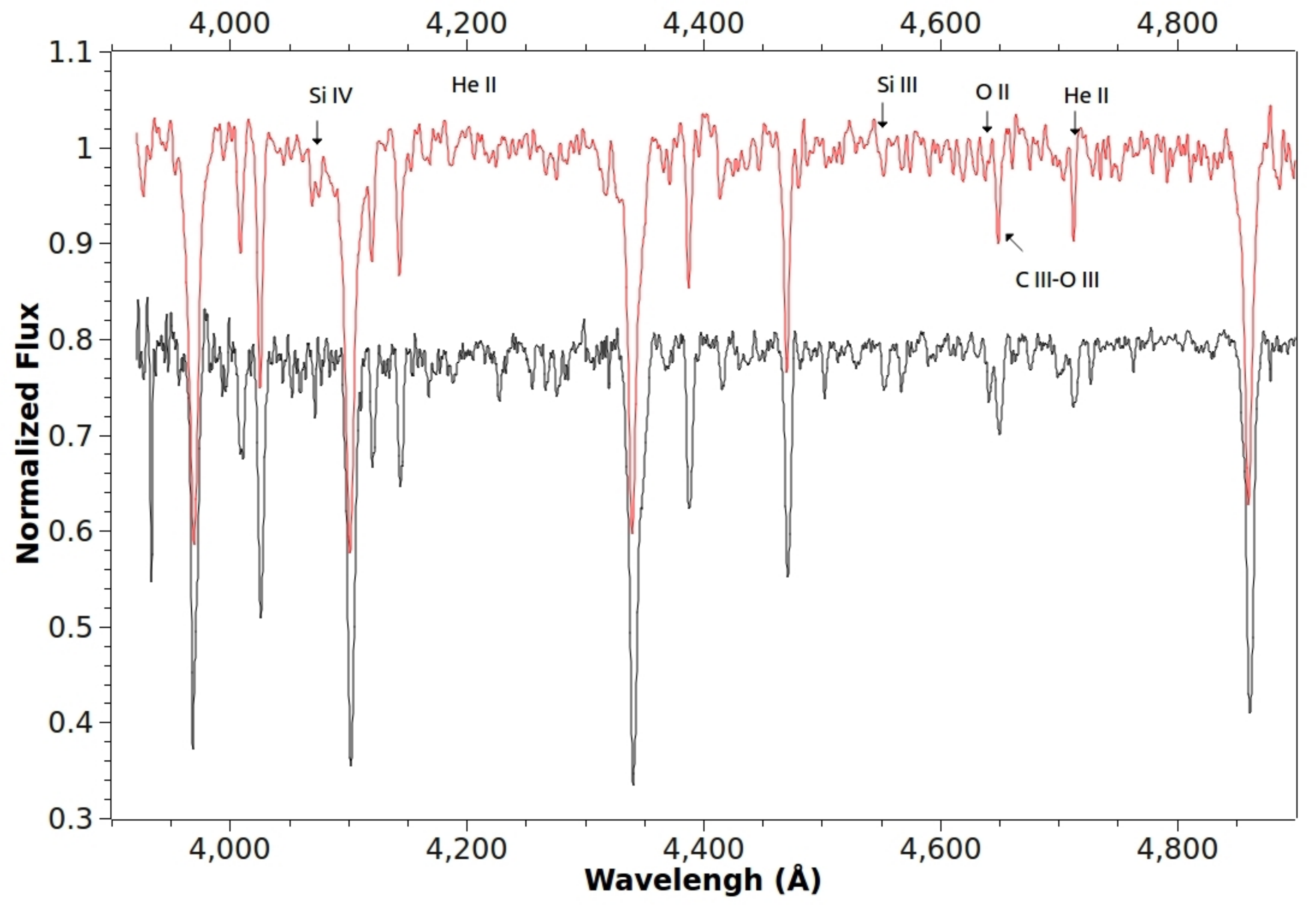}
\caption{
The FLECHAS spectrum of HD\,36665 (black) in comparison with a B1V template.
}
\label{f:hd36665}
\end{figure}

The sources observed by CAFOS are foreground stars with a maximum distance of $671\pmpm{+96}{-56}$ \pc.
Their spectral properties were identified by the same procedure as done for HD\,37424.
However, due to the lower resolution CAFOS spectra have higher uncertainty (Table \ref{t:obs_result}).
We used FLECHAS to identify the spectral types of the background star HD\,36665.
This is a B1$\pm$0.5V type star (Figure \ref{f:hd36665}).

\begin{table}
\caption{Observed stars in the region.
Names, angular separations to the SNR GC (in arcmin), spectral types, instruments used and
distances with upper and lower errors (in \pc) are given.
The error in temperature subclasses is $\pm$1 for those with integer value and $\pm$0.5 for those with half integer value.
The resolution of CAFOS is not enough to distinguish the luminosity classes IV and V.
Yet, the stars are assumed to be in luminosity class V.}
\label{t:obs_result}
\begin{center}
\begin{tabular}{@{}c@{~~}c@{~~}c@{~~}c@{~~}c@{~}r@{~/~}l@{}}
Name
& Angular
& Spec.
& Inst.$^1$
& Distance
& \multicolumn{2}{c}{Distance} \\

& Separation
& Type
&
&
& \multicolumn{2}{c}{(Error)} \\\hline
TYC 1869-01281-1 & 19.4 & A2V   & C   & 574 & +73& -41  \\
TYC 1869-01317-1 & 5.5  & B9.5V & T   &1013 &+183&-174  \\
TYC 1869-01334-1 & 25.5 & F2V   & C   & 289 & +38& -34  \\
TYC 1869-01376-1 & 27.9 & A7V   & C   & 428 & +18& -32  \\
TYC 1869-01505-1 & 23.5 & F1V   & C   & 380 & +50& -44  \\
TYC 1869-01610-1 & 17.6 & F7V   & C   & 245 & +18& -16  \\
TYC 1869-01632-1 & 27.2 & F8V   & C   & 316 & +37& -26  \\
TYC 1869-01642-1 & 5.61 & B9.5V & T   & 952 &+172&-163  \\
TYC 1869-01679-1 & 25.6 & G2V   & C   & 170 & +11&  -9  \\
TYC 1869-01749-1 & 20.1 & F8V   & C   & 166 & +19& -14  \\
TYC 1873-00145-1 & 25.0 & A3V   & C   & 603 & +46& -50  \\
TYC 1873-00307-1 & 29.2 & F8V   & C   & 326 & +40& -29  \\
TYC 1873-00347-1 & 19.8 & A0V   & C   & 671 & +96& -56  \\
UCAC4-589-020390 & 22.1 & F4V   & C   & 381 & +14& -16  \\\hline
\multicolumn{7}{l}{$^1$ Instruments: C: CAFOS, T: TFOSC} \\
\end{tabular}
\end{center}
\end{table}

TRES data were used for radial velocity (RV) measurements of HD\,37424.
Due to the low signal to noise, 10 absorption features could be used (Table \ref{t:rv}).
Doublets or triplet lines are avoided as these features increased the dispersion significantly.
The lines were fitted by Gaussian functions and the best shifts were determined.
Two more Gaussian fits with different centers were applied for each line;
one fits the outer edge of the noise of the blue side while fitting the inner edge of the red.
The other fits the outer edge of the red and the inner edge of the blue.
By doing this, underestimating the errors is avoided.
As the features are wide and the data are noisy,
the centers of these secondary fits are far away from the best fits, 16 \kms on average.
The difference between the best fit and the upper limit (towards red) fit is distinct from
the best fit and the lower limit (towards blue) difference.
To avoid underestimating the error, the greater difference is assigned as the final error.
The average heliocentric velocities of the first and the second spectra are -9.2 and -9.1 \kms
with line to line scattering standard deviations being 6.5 and 5.5 \kms respectively.

\begin{table}
\caption{Radial velocity measurements of HD\,37424.
The laboratory wavelength of the features, the Gaussian width ($\sigma$) of the fit,
the wavelength shift obtained by the best fitting Gaussian,
upper and lower limits of the shift from different Gaussian fits and the final error are presented.
The features are given in \AA\ while all other columns are in \kms}
\label{t:rv}
\begin{center}
\begin{tabular}{@{}c c c c c c@{}}
& & \multicolumn{4}{c}{Observations on September 16th, 2013} \\\cline{3-6}
\noalign{\smallskip}
Feature   & $\sigma$  & BF	   & UL		 & LL		 & E	\\\hline
\noalign{\smallskip}
3889.051  & 164	   & -10.1	& 19.3	   & -29.2	  & 29.4	\\
3970.074  & 137	   & -0.5	 & 21.4	   & -22.3	  & 21.9	\\
4101.737  & 144	   & -9.4	 & 12.2	   & -32.3	  & 22.9	\\
4143.759  & 98		& +4.0	 & 17.1	   & -10.9	  & 14.9	 \\
4340.468  & 142	   & -10.6	& 6.2		& -28.8	  & 18.2	\\
4387.928  & 81		& -13.1	& -7.1	   & -24.1	  & 11	\\
4861.332  & 139	   & -13.8	& 0.9		& -18.9	  & 14.7   \\
4921.929  & 89		& -16.5	& -1.6	   & -32.4	  & 15.9	\\
6562.817  & 115	   & -14.7	& 0.3		& -21.5	  & 15	 \\
6678.149  & 82		& -7.5	 & 0.3		& -14.5	  & 7.8	 \\\hline
\noalign{\smallskip}
& & \multicolumn{4}{c}{Observations on September 21st, 2013} \\\cline{3-6}
\noalign{\smallskip}
Feature   & $\sigma$  & BF	   & UL		 & LL		 & E	\\\hline
\noalign{\smallskip}
3889.051  & 164	   & -12.6	& 6.2		& -28.2	  & 18.8	 \\
3970.074  & 137	   & -1.4	 & 18.0	   & -19.8	  & 19.4	\\
4101.737  & 144	   & -5.1	 & 14.9	   & -21.2	  & 20	 \\
4143.759  & 98		& -1.9	 & 7.2		& -8.7	   & 9.1	\\
4340.468  & 142	   & -11.5	& 11.9	   & -36.6	  & 25.1	 \\
4387.928  & 81		& -8.8	 & -4.1	   & -19.3	  & 10.5	\\
4861.332  & 139	   & -11.7	& -3.0	   & -19.5	  & 8.7	  \\
4921.929  & 89		& -17.2	& -4.9	   & -31.6	  & 14.4	\\
6562.817  & 115	   & -14.4	& 1.4		& -29.7	  & 15.8	\\
6678.149  & 82		& -6.7	 & 3.3		& -15.1	  & 10	\\\hline
\noalign{\smallskip}
\multicolumn{6}{l}{BF: Best fit, UL: Upper Limit, LL: Lower Limit, E: Error}\\
\end{tabular}
\end{center}
\end{table}

\begin{figure*}
\includegraphics[width=5.5cm]{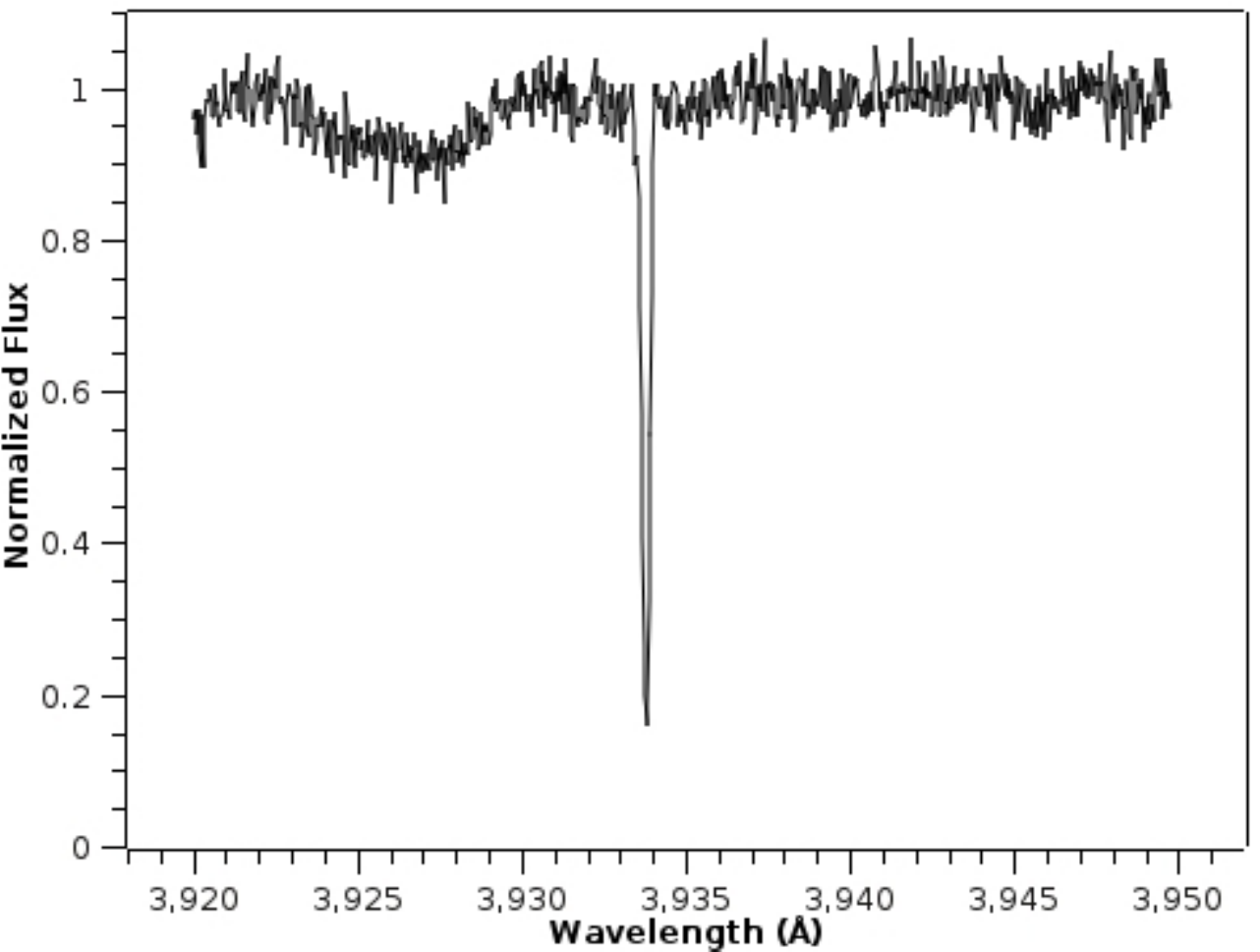}
\includegraphics[width=5.5cm]{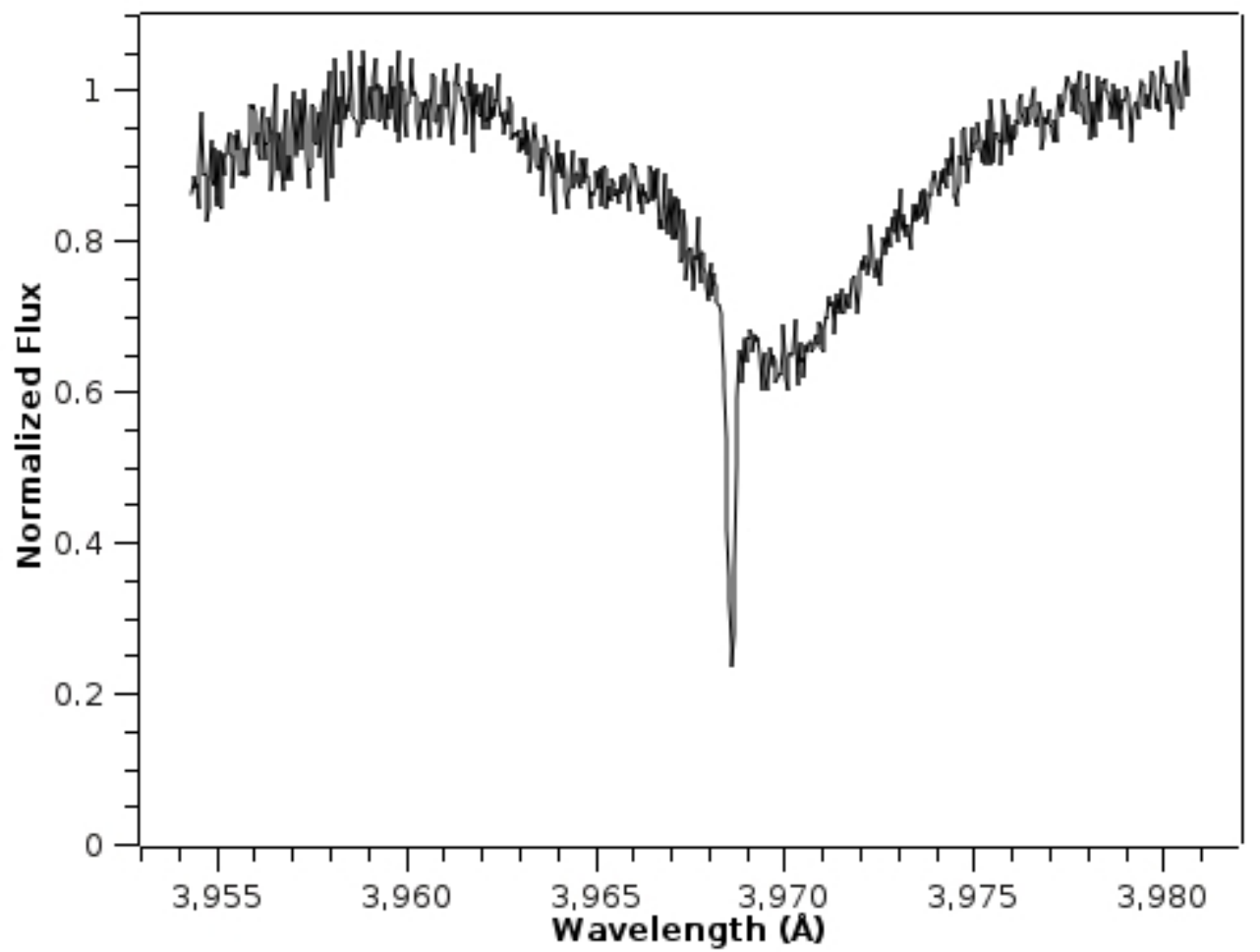}
\includegraphics[width=5.5cm]{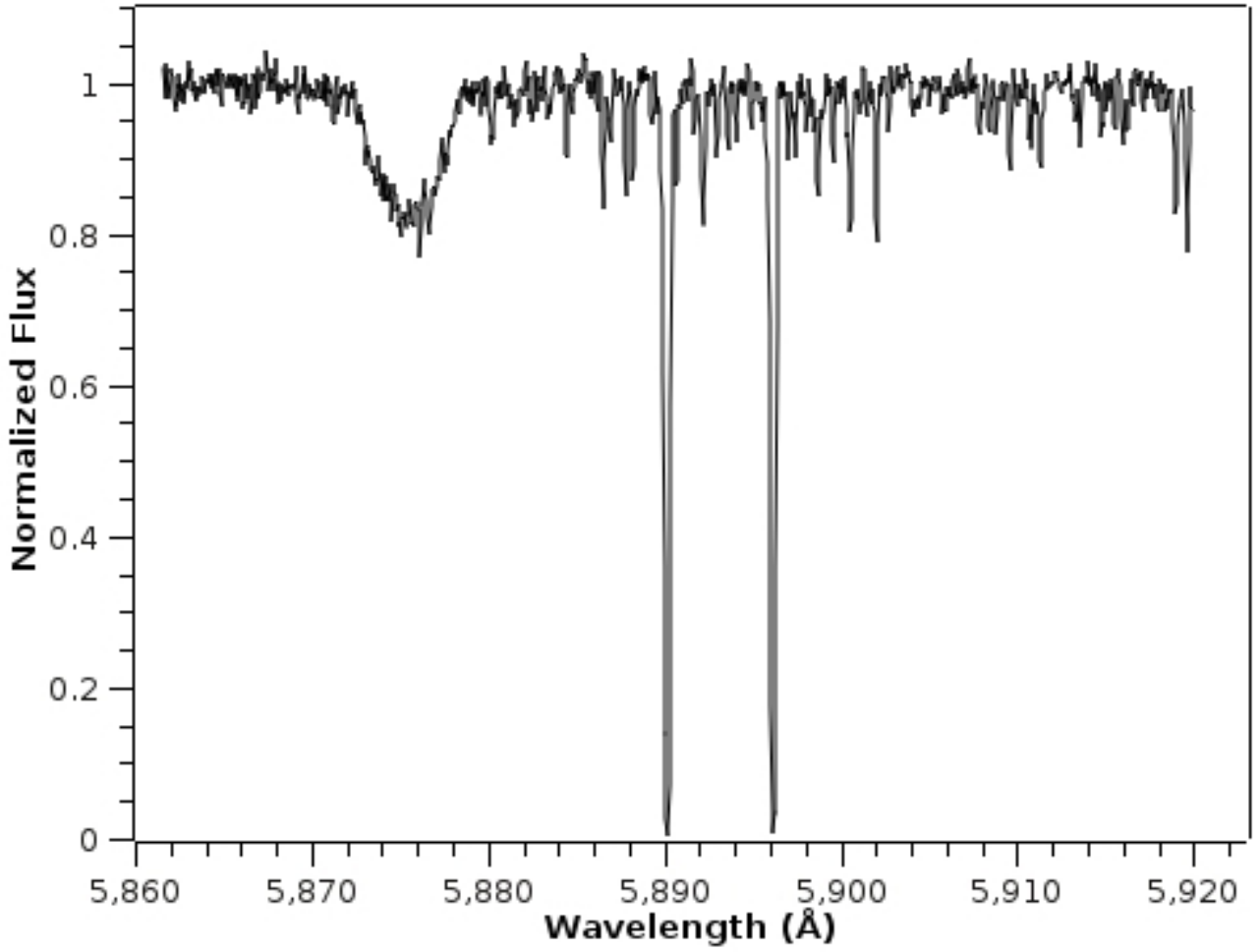}
\caption{The strong interstellar lines of HD37424.
\textit{Left}: \ion{Ca}{ii}--K line from TRES spectrum.
Despite a weak blended feature at -13 \kms, there is no clear high velocity component.
\textit{Middle}: \ion{Ca}{ii}--H line (H$\epsilon$ as a background) from the same spectrum.
Again there is no clear high velocity gas component.
\textit{Right}: Interstellar \ion{Na}{i}--D1 and D2 lines.
The feature with high FWHM is the \ion{He}{i} $\lambda$5875 triplet. Those around Sodium doublet are tellurics.
}
\label{f:3933}
\end{figure*}

As long as the star is inside the supernova remnant,
high velocity gas accelerated by the SNR is expected to reveal itself as
blue--shifted components to \ion{Ca}{ii}--K and H and/or \ion{Na}{i}--D1 and D2 absorption features.
The spectra show no clear high velocity features (Figure \ref{f:3933}).
The average heliocentric velocity of the interstellar gas is 12.1$\pm$0.5 \kms (Table \ref{t:ca_vel}).
This is typical for the neighboring stars
\citep{1973ApJ...181..799S}.
The star has no positional correspondence with bright filaments, but with fainter \halpha emitting regions.
It is located at the vicinity of the eastern cavity seen in the radio and $\gamma$--ray images
\citep{1980A&A....92..225K,
	   2012ApJ...752..135K}.
As the star can hardly be a foreground source,
we suggest that, in this direction, the shocked ejecta has not reached the dense interstellar gas yet.
The reader must note that, the background stars displaying high velocity \ion{Ca}{ii} lines in
\citet{2004A&A...426..555S}
are located behind bright filament knots.
The high velocity gas is found where the filaments are concentrated
\citep{1976SvA....20...19L}.

\begin{table}
\caption{Measured velocities for interstellar \ion{Ca}{ii}--K and H and \ion{Na}{i}--D1 and D2 lines.
These lines have a low Gaussian width unlike the intrinsic features of the star.
The average velocity is 12.1 \kms with a standard deviation 0.5 \kms in each observation.
The data in the first column are in (\AA) while the others are in \kms.}
\label{t:ca_vel}
\begin{center}
\begin{tabular}{c c c c}
Feature   & $\sigma$  & Velocity & Velocity \\
 &  & (Day 1) & (Day 2) \\\hline
\noalign{\smallskip}
3933.664  & 7.5	   & 12.3   & 11.8  \\
3968.47   & 5.9	   & 12.8   & 12.7  \\
5889.953  & 9.0	   & 11.6   & 11.6  \\
5895.923  & 8.3	   & 11.9   & 12.0  \\\hline
\noalign{\smallskip}
\multicolumn{4}{l}{Day 1: Sep. 16th, 2013; Day 2: Sep 21st, 2013.}\\
\end{tabular}
\end{center}
\end{table}

Using the absolute visual magnitude from \citet{1982lbg6.conf.....A},
for a spectral type B0.5V$\pm$0.5, the distance modulus yields $1868\pmpm{+410}{-403}$ \pc for HD\,37424.
Using the luminosity of the same spectral type from
\citet{2010AN....331..349H}, we find the distance as 1318$\pm$119 \pc which is well consistent with the distance of the pulsar.
The total visual absorption was taken into account in both calculations.

As the absolute magnitudes for early type stars can range from star to star in the same temperature,
interstellar \ion{Ca}{ii} lines are also used in distance determination.
By using the following relation from \citet{2009A&A...507..833M},
\begin{equation}
D = 77+\left(2.78+\frac{2.60}{\frac{EW(K)}{EW(H)}-0.932}\right)EW(H)
\end{equation}
the distance to HD\,37424 is found to be $1288\pmpm{+304}{-193}$ \pc which is almost in the same range as the pulsar.
The equivalent widths are 243$\pm$7 and 160$\pm$15 m\AA\ for \ion{Ca}{ii}--K and \ion{Ca}{ii}--H lines respectively.

The extinction towards the star was also derived from 4430 and 4502 \AA\ DIB's.
The measured equivalent widths have high error due to the blending.
Using the EW's, the E(B-V) was calculated based on the relation mentioned in
\citet{1975ApJ...196..129H}.
It yields 0.42$\pm$0.08 mag; points to somewhat larger color excess but does not exclude 0.35$\pm$0.04 mag derived from photometry.
\section{Kinematics}

HD\,37424 and PSR\,J0538+2817 have proper motions receding from each other and departing from the same location on the sky (Figure \ref{f:mosaic}).
The proper motion of both objects are corrected for Galactic rotation and solar motion.
The Galactocentric distance to the Sun is taken as 8.5 \kpc and the solar rotational velocity as 220 \kms.
The local standard of rest was taken from \citet{2011MNRAS.410..190T} as
(U$_{\odot}$,V$_{\odot}$,W$_{\odot}$) =
(10.4$\pm$0.4, 11.6$\pm$0.2, 6.1$\pm$0.2) \kms.
The resultant values for HD\,37424 are
$\mu_{\alpha}^*=10.0\pm0.8$ \masyr,
$\mu_{\delta}=-5.9\pm0.6$ \masyr and
for the pulsar,
$\mu_{\alpha}^*=-24.4\pm0.1$ \masyr,
$\mu_{\delta}=57.2\pm0.1$ \masyr.
Together with the $-20.0\pm6.5$ \kms peculiar radial velocity, HD\,37424 has a space velocity of 74$\pm$8 \kms at 1.3 \kpc.
HD\,37424 is a runaway star of which velocity is higher than typical runaway velocities 40--50 \kms.
The 2--D space velocity of the pulsar at the same distance is 382.2$\pm$0.8 \kms.

We constructed the past 3D trajectories of PSR\,J0538+2817 and the runaway star HD\,37424
to evaluate whether these two objects could have been at the same place at the same time in the past.
Since we applied the same method already in preceding papers \citep{2010MNRAS.402.2369T,2011MNRAS.417..617T,2012PASA...29...98T},
we refer to these publications for details.
We construct three million past trajectories of PSR\,J0538+2817 and HD\,37424
throughout Monte Carlo simulations by varying the observables (parallax, proper motion, radial velocity) within their error intervals.
For the radial velocity of the NS, we assume a uniform distribution in the range of -1500 to +1500 \kms.
From all pairs of trajectories, we evaluate the smallest separation $d_{min}$ and the past time $\tau$ at which it occurred.
The distribution of separations $d_{min}$ is supposed to obey the distribution of absolute differences of
two 3D Gaussians (see e.\,g. \citealt{2001A&A...365...49H}, equations A3 and A4; \citealt{2012PASA...29...98T}, equations 1 and 2),
if it is assumed that the stellar 3D positions are Gaussian distributed.
Since the actual (observed) case is different from this simple model
(no 3D Gaussian distributed positions, due to e.\,g. the Gaussian distributed parallax that goes into the position reciprocally,
uniform radial velocity distribution of the NS, etc.), we adapt the theoretical formulae (we use equations 1 and 2 in
\cite{2012PASA...29...98T}
with the symbols $\mu$ and $\sigma$ for the expectation value and standard deviation, respectively)
only to the first part of the $d_{min}$ distribution (up to the peak plus a few more bins, see \citealt{2012PASA...29...98T}).
The derived parameter $\mu$ then gives the positional difference between the two objects.
\footnote{The uncertainties on the separation are dominated by the kinematic uncertainties of the NS
that are typically of the order of a few hundred \kms  (because of the assumed radial velocity distribution).
As a consequence, the distribution of separations  $d_{min}$ shows a large tail for larger separations.
However, the first part of the $d_{min}$ distribution (slope and peak) can still be explained well with the theoretical curve
(here equation 2 in \citealt{2012PASA...29...98T}) since the kinematic dispersion for only those runs are much smaller, a few tens of \kms for the NS, i.e. a few tens of pc after 1 Myr.}

\begin{figure}
\includegraphics[width= 8 cm]{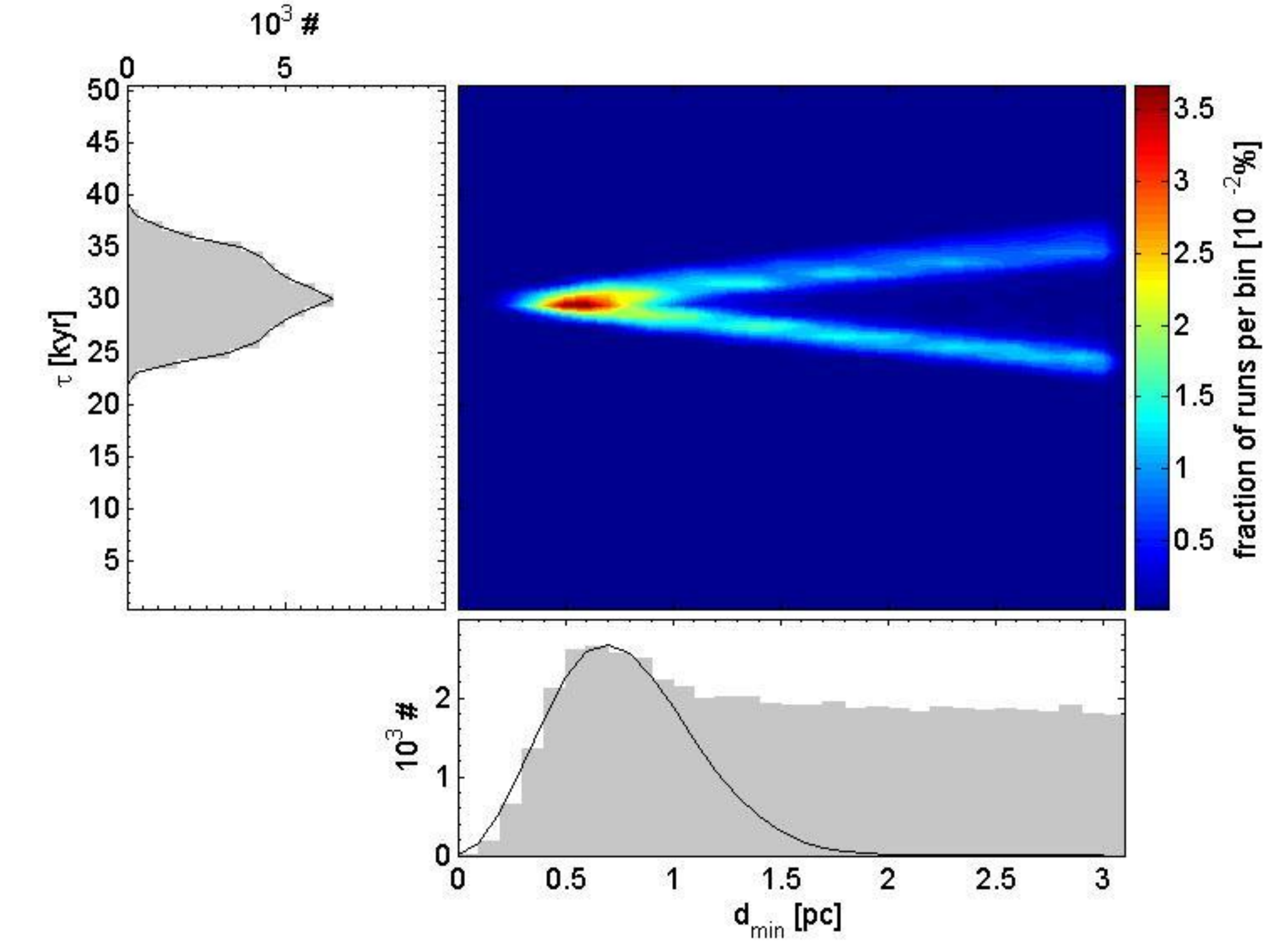}
\caption{Distributions of minimum separations $d_{min}$ and corresponding flight times $\tau$ for encounters between PSR\,J0538+2817
and HD\,37424. The solid curves drawn in the $d_{min}$ histogram (bottom panel) represent the theoretically expected distribution with $\mu=0$ and $\sigma=0.35$ \pc, adapted to the first part of the histogram.}
\label{f:nina}
\end{figure}

We find that both stars were at the same position in the past
\ie
$\mu=0$ at
$\left(l, b\right)=\left(84.82\pm0.01, 27.84\pm0.01\right)$ deg at
$30\pm4$ \kpc (Figure \ref{f:nina}).
This predicted position of the supernova is $4.2\pmpm{+0.8}{-0.6}$ arcmin offset from the nominal geometric center.
The predicted distance of the supernova (as it is seen from the Earth today) is $1333\pmpm{+103}{-112}$ \pc.
\section{Associations}

There is no known OB association within 4.5$^{\circ}$ ($\sim$100 \pc at 1300 \pc) from SNR S147.
Considering the angular separations and radial distances, the progenitor star cannot be linked to any of the young open cluster around.
Hence, it is considered to be a runaway star ejected \eg from the cluster NGC 1960, which is 217 \pc away from the SNR, several millions years ago
\citep{2007ApJ...654..487N}.
However, binarity among cluster ejected runaway stars are rare
\citep{1986ApJS...61..419G}.
So, as the progenitor was the previous binary companion to HD\,37424, it may be unlikely that they were ejected from a cluster.
Therefore, the neighboring stars are investigated to search for any association.
All of the OB type stars within 4.5$^{\circ}$ (100 pc at 1.3 kpc distance) of the geometrical center (GC) of the SNR
were selected from \textit{The Catalogue of Stellar Spectral Classifications}
\citep{2013yCat....1.2023S}.
The spectral types are known through spectroscopic observations.
There are some discrepancies in spectral types reported by different papers.
So, always the latest reference was chosen.
Together with 14 stars from our observations at TUG and Calar Alto, 99 stars are used.
The photometric data were obtained from the catalog \textit{ASCC-2.5 V3}
\citep{2001KFNT...17..409K} except for UCAC4--589--020390 of which photometric
data was retrieved from the \textit{UCAC4} catalog.

For each star having an integer spectral subclass, distance and extinction are calculated in an interval
between one spectral subclass above and below, \eg B0V--B2V for a B1V type star.
For those having half integer subclass, half spectral subclass above and below are used \eg B0V--B1V for a B0.5V type star.
The total visual extinction, \Av, was determined by using BVJHK colors and intrinsic color differences of corresponding spectral types.
Using the following relation, \Av values were derived for each color difference; B-J, B-H, B-K, V-J, V-H, V-K.
\begin{equation}
\Av =
\frac{(\lambda_{1}-\lambda_{2}) - (\lambda_1-\lambda_2)_0}%
{\mathrm{A_{\lambda_1}}/\Av-\mathrm{A_{\lambda_2}}/\Av},
\end{equation}
The intrinsic colors are obtained from
\citet{1995ApJS..101..117K} and
\citet{1994MNRAS.270..229W},
while $\mathrm{A_{\lambda}}$/$\mathrm{A_{V}}$ ratios are from
\citet{1985ApJ...288..618R}.
Color differences of short intervals, \ie H-K have high errors as they are multiplied by higher coefficients.
The color excess E(B-V) is mentioned separately.
(Table \ref{t:26_stars})
\Av values obtained from six color differences were averaged and
their standard deviation were calculated.
This was applied for all three possible spectral types of the source.
As long as the error due to the spectral type uncertainty is larger than the standard deviation of the color differences,
it was assigned to be the final error.
When the error is asymmetric due to the intrinsic colors of different spectral types, the larger one was accepted.
In some cases, the uncertainty is dominated by the error in color differences.
Then, these were preferred to be the final error for such sources.
E(B-V) versus \Av fit
yields a total to selective absorption ratio is 3.24$\pm$0.06.
In individual cases, E(B-V) deviates strongly from \Av.
However, the large sample reveals that the ratio of total to selective absorption has a usual value.
The distances were derived from distance moduli using the absolute magnitudes from
\citet{1982lbg6.conf.....A}.
Errors in distances are due to the uncertainty in spectral type and the error in $\mathrm{A_{V}}$ were also taken into account.

The spectro--photometric distance alone is not enough
due to the high dispersion in the brightness of OB--type stars
\citep{2006MNRAS.371..185W} (Thereafter W06).
Assuming the stars beyond 1 \kpc are within 100 \pc from the SNR GC, the absolute visual magnitudes were calculated.
Although the dispersion mentioned in W06 is high, 24 of the stars fit well in the comparison with the W06 values.
Hence, we suggest that these stars are members of an OB association.
(Table \ref{t:26_stars}, Table \ref{t:M_stars},
19 of them have similar pm values.
The average pm in right ascension is -1.39 \masyr and -4.17 \masyr
in declination with 0.99 \masyr and 1.45 \masyr standard deviation respectively (Table \ref{t:13_stars}).
At 1.3 \kpc, 5 of them are runaway stars exceeding 20 \kms 2--D peculiar velocity.
This is consistent with the general ratio of the runaway stars to the normal stars which is 10--30 percent
\citep{1987ApJS...64..545G}.
\begin{table*}
\caption{24 stars beyond 1 \kpc are presented. Angular distances are given in degrees, distances in parsecs, extinctions and brightness in visual band in magnitudes.
In the column; SpT adopted, the average spectral types that are used in distance calibrations are given.
}
\label{t:26_stars}
\begin{center}
\begin{tabular}{@{}c@{~~~}c@{~~~}c@{~~~}c@{~~~}c@{~~~}c@{~~~}c@{~~~}c@{~~~}c@{~~}r@{~/~}lc@{~~~}c@{~~~}c@{~~~}c@{}}
Ang. Sep.
& Name
& SpT
& SpT
& $\mathrm{A_{V}}$
& $\mathrm{A_{V}}$
& E(B-V)
& E(B-V)
& Distance
& \multicolumn{2}{c}{Distance}
& V
& Ref\#$^*$ \\

&
&
& (adopted)
&
& (Err)
&
& (Err)
&
& \multicolumn{2}{c}{(Err)}
&
&  \\\hline
0.09 & TYC 1869-01317-1 & B9.5V        & B9.5V      & 0.83  & 0.1  & 0.26 & 0.090 & 1013 &  +183&-174   &  11.258 &  tw   \\
0.09 & TYC 1869-01642-1 & B9.5V        & B9.5V      & 1.00  & 0.1  & 0.3  & 0.090 & 952  &  +172&-163   &  11.293 &  tw   \\
0.17 & HD 37424         & B0.5V        & B0.5V      & 1.28  & 0.06 & 0.35 & 0.036 & 1868 &  +410&-403   &  8.989 &  tw   \\
0.52 & HD 36993         & B0.5/1III/IV & B0.5III-IV & 1.35  & 0.06 & 0.39 & 0.028 & 1961 &  +636&-628   &  8.248 &  1    \\
0.63 & HD 37318         & B0.5Ve       & B0.5V      & 2.29  & 0.15 & 0.59 & 0.026 & 903  &  +231&-228   &  8.399 &  1    \\
1.00 & BD+27 797        & B0.5Ve       & B0.5V      & 2.50  & 0.14 & 0.62 & 0.094 & 1788 &  +485&-465   &  10.095 &  2    \\
1.35 & HD 37696         & B0.5IV/V     & B0.5IV-V   & 1.12  & 0.06 & 0.29 & 0.023 & 1549 &  +565&-558   &  7.973 &  1    \\
1.35 & BD+27 850        & B1.5IVe      & B1.5V      & 1.31  & 0.04 & 0.35 & 0.056 & 1525 &  +313&-305   &  9.362 &  2    \\
1.52 & HD 245770        & O9/B0III/Ve  & B0III-V    & 2.11  & 0.13 & 0.79 & 0.053 & 2595 &  +1090&-1066 &  9.187 &  3    \\
1.76 & HD 36441         & B0.5/1.5V    & B1V        & 1.13  & 0.13 & 0.33 & 0.058 & 1153 &  +628&-389   &  8.239 &  1    \\
1.97 & BD+26 943        & B2V          & B1.5V      & 1.41  & 0.14 & 0.37 & 0.040 & 1357 &  +723&-511   &  9.623 &  8   \\
2.61 & HD 38010         & B1III        & B1III      & 1.29  & 0.1  & 0.22 & 0.028 & 967  &  +437&-236   &  6.817 &  4    \\
2.95 & BD+30 938        & B3III        & B3III      & 1.44  & 0.04 & 0.45 & 0.030 & 1332 &  +741&-255   &  9.062 &  7    \\
2.98 & Sh 2-242 1       & B0V          & B0V        & 2.17  & 0.13 & 0.78 & 0.056 & 2318 &  +853&-842   &  9.996 &  5    \\
3.06 & HD 37366         & O9.5V        & O9.5V      & 1.22  & 0.06 & 0.35 & 0.016 & 1375 &  +201&-199   &  7.640 &  6    \\
3.07 & BD+30 987        & B5III:       & B5III      & 0.73  & 0.08 & 0.25 & 0.044 & 1523 &  +398&-287   &  9.444 &  7    \\
3.24 & BD+30 976        & B7V          & B7V        & 0.89  & 0.17 & 0.18 & 0.020 & 994  &  +264&-226   &  10.277 &  7    \\
3.45 & BD+25 989        & B1Vn         & B1V        & 1.59  & 0.13 & 0.43 & 0.082 & 1872 &  +1061&-649  &  9.751 &  8    \\
3.61 & BD+27 909        & B2III        & B2III      & 1.87  & 0.1  & 0.55 & 0.076 & 2004 &  +683&-761   &  9.479 &  8    \\
4.05 & BD+31 1065       & B3III        & B3III      & 0.67  & 0.04 & 0.27 & 0.043 & 2304 &  +1286&-442  &  9.482 &  7    \\
4.07 & BD+31 1050       & B3III        & B3III      & 1.01  & 0.03 & 0.35 & 0.047 & 1791 &  +988&-338   &  9.276 &  7    \\
4.09 & HD 38909         & B3II-III      & B3II-III    & 0.56  & 0.03 & 0.17 & 0.033 & 2057 &  +1007&-992  &  8.145   &  9    \\
4.28 & BD+31 1021       & B7V          & B7V        & 1.05  & 0.12 & 0.14 & 0.106 & 1007 &  +250&-218   &  10.465  &  7    \\
4.46 & HD 40297         & B9.5Ib/II    & B9.5Ib/II  & 1.07  & 0.13 & 0.25 & 0.015 & 1337 &  +693&-688   &  7.270    &  1    \\\hline
\multicolumn{13}{p{15cm}}{%
$^*$ Ref\#: %
1:  \cite{1979RA......9..479C}; %
2:  \cite{1999A+AS..137..147S}; %
3:  \cite{1998PASP..110.1310W}; %
4:  \cite{1976A+AS...26..241C}; %
5:  \cite{1990AJ.....99..846H}; %
6:  \cite{1971ApJS...23..257W}; %
7:  \cite{1961AnTou..28...33B}; %
8:  \cite{1977ApJ...217..127C}; %
9:  \cite{1955ApJS....2...41M}; %
10: \cite{1977ApJ...217..127C}; %
tw: This work.} \\
\end{tabular}
\end{center}
\end{table*}
\begin{table}
\caption{Visual absolute magnitudes at 1.3$\pm$0.1 \kpc for the stars of the possible OB association of which HD 37424 is a member.
Errors are mainly due to the distance range 1.2$-$1.4 \kpc.
The expected M$_{V}$ for the corresponding spectral types and its dispersion from
\citet{2006MNRAS.371..185W} (W06)
are given. All values are in mag.}
\label{t:M_stars}
\begin{center}
\begin{tabular}{@{}c@{~~}c@{~}r@{~/~}l@{~~}c@{~~}c@{}}

& $\mathrm{M_{V}}$
& \multicolumn{2}{c}{error}
& $\mathrm{M_{V}}$
& Disp. \\
Name
& (@1.3$\pm$0.1 \kpc)
& \multicolumn{2}{c}{~}
& (W06)
& (W06) \\\hline
TYC 1869-01317-1 &  -0.14 &  +0.26&-0.27  & +0.29   & 1.40 \\
TYC 1869-01642-1 &  -0.28 &  +0.26&-0.27  & +0.29   & 1.40 \\
HD 37424         &  -2.86 &  +0.22&-0.23  & -3.34   & 2.40 \\
HD 36993         &  -3.67 &  +0.22&-0.23  & -4.02   & 2.30 \\
HD 37318         &  -4.46 &  +0.31&-0.32  & -3.34   & 2.40 \\
BD+27 797        &  -2.97 &  +0.30&-0.31  & -3.34   & 2.40 \\
HD 37696         &  -3.72 &  +0.22&-0.23  & -3.34   & 2.40 \\
BD+27 850        &  -2.52 &  +0.20&-0.21  & -2.95   & 1.16 \\
HD 245770        &  -3.49 &  +0.29&-0.30  & -3.34   & 2.40 \\
HD 36441         &  -3.46 &  +0.29&-0.30  & -2.95   & 1.16 \\
BD+26 943        &  -2.36 &  +0.30&-0.31  & -2.64   & 1.40 \\
HD 38010         &  -5.04 &  +0.26&-0.27  & -4.10   & 2.20 \\
BD+30 938        &  -2.95 &  +0.20&-0.21  & -2.32   & 1.50 \\
Sh 2-242 1       &  -2.74 &  +0.29&-0.30  & -3.34   & 2.40 \\
HD 37366         &  -4.15 &  +0.22&-0.23  & -4.49   & 2.27 \\
BD+30 987        &  -1.86 &  +0.24&-0.25  & -1.49   & 2.00 \\
BD+30 976        &  -1.18 &  +0.33&-0.34  & -0.63   & 1.40 \\
BD+25 989        &  -2.41 &  +0.29&-0.30  & -2.95   & 1.16 \\
BD+27 909        &  -2.96 &  +0.26&-0.27  & -2.63   & 2.20 \\
BD+31 1065       &  -1.76 &  +0.20&-0.21  & -2.32   & 1.50 \\
BD+31 1050       &  -2.30 &  +0.19&-0.20  & -2.32   & 1.50 \\
HD 38909         &  -2.98 &  +0.19&-0.20  & -2.32   & 1.50 \\
BD+31 1021       &  -1.15 &  +0.28&-0.29  & -0.63   & 1.40 \\
HD 40297         &  -4.37 &  +0.29&-0.30  & -3.75   & 1.30 \\\hline
\end{tabular}
\end{center}
\end{table}
\begin{table}
\caption{19 stars with common proper motion are presented.
The average PM$_{\text{RA}}$ is -1.39 \masyr and PM$_{\text{Dec}}$ is -4.17 \masyr.
The last three column represents the 2--D space velocity of
the stars with respect to the average motion of the group with maximum and minimum values in \kms.
All the values of proper motions are in \masyr.}
\label{t:13_stars}
\begin{center}
\begin{tabular}{c c c c c c c c}
Name              &  $\mu_{\alpha}^{*}$     & err     & $\mu_{\delta}$      & err     & $\mathrm{V_{REL}}$      & max      & min \\\hline
HD 37318          &  -0.8 &  0.6 & -5.9  & 0.6 &  11.2 & 16.1 & 6.9 \\
TYC 1869-01642-1  &  -1.5 &  1.1 & -2.4  & 1.2 &  11.0 & 19.8 & 7.1 \\
HD 38010          &  -1.7 &  1.0 & -3.3  & 1.0 &  5.7  & 14.1 & 0.0 \\
BD+30 976         &  -3.5 &  0.9 & -5.3  & 0.5 &  14.7 & 21.1 & 8.4 \\
BD+31 1021        &  -2.8 &  0.7 & -5.5  & 0.8 &  11.9 & 18.4 & 5.4 \\
TYC 1869-01317-1  &  -0.9 &  0.9 & -5.1  & 0.9 &  6.5  & 14.2 & 0.0 \\
HD 36441          &  -2.0 &  0.6 & -6.6  & 0.6 &  15.4 & 20.1 & 11.2\\
BD+30 938         &  -1.7 &  0.8 & -3.6  & 0.6 &  4.0  & 9.9  & 0.0 \\
HD 40297          &  -1.4 &  1.0 & -3.7  & 1.0 &  2.9  & 11.0 & 0.0 \\
BD+26 943         &  -0.9 &  0.8 & -3.7  & 0.8 &  4.2  & 11.2 & 0.0 \\
BD+30 987         &  -2.0 &  0.6 & -4.7  & 1.2 &  4.9  & 13.0 & 0.0 \\
BD+27 797         &  -0.1 &  0.7 & -3.1  & 1.4 &  10.4 & 19.6 & 4.2 \\
BD+31 1050        &  -1.0 &  0.6 & -4.0  & 0.6 &  2.7  & 7.8  & 0.0 \\
BD+25 989         &  -1.5 &  0.9 & -2.5  & 1.6 &  10.3 & 21.1 & 4.9 \\
HD 36993          &  -0.2 &  0.9 & -5.8  & 0.6 &  12.4 & 18.8 & 6.6 \\
HD 38909          &  +0.7 &  0.5 & -4.2  & 1.5 &  12.9 & 18.4 & 0.0 \\
BD+31 1065        &  -1.0 &  0.6 & -1.3  & 0.5 &  17.9 & 21.7 & 14.7\\
Sh 2-242 1        &  -0.1 &  0.7 & -1.8  & 1.0 &  16.7 & 24.2 & 9.2 \\
HD 245770         &  -2.0 &  0.5 & -4.3  & 1.1 &  3.8  & 10.2 & 0.0 \\\hline
\end{tabular}
\end{center}
\end{table}
\section{Discussion}

The runaway nature of HD\,37424 is clear.
The chance projection of such a massive runaway star moving away from the GC of an SNR in the Galactic anti--center direction must be very low.
In addition, combining with the central compact object after tracing back both objects at the same time
shows that HD\,37424 is clearly the pre--supernova binary companion of the progenitor of SNR S147 and PSR\,J0538+2817.
BSS is the favored explanation for its runaway nature.

HD\,37424 is a B0.5V type star with a mass of $\sim$13 \msun
\citep{2010AN....331..349H}.
So, the progenitor of the pulsar must have a higher mass.
Based on the lack of O type stars in the field (see Table \ref{t:26_stars})
we set an upper mass limit of 20--25 \msun.
It may even imply a twin binary.
The Roche Lobe radii calculated for 15, 20 and 25 \msun vary between 91 and 311 \rsun which shows that the system might have been an interacting binary.
Hence, the progenitor star should be a naked helium star at the final stage of its evolution with a mass even as low as 2 \msun
\citep{1993SSRv...66..309V},
\citep{1995ApJ...448..315W}.
However, how conservative the mass transfer was, will be understood after further observations.
Assuming a circular orbit, pre--supernova binary parameters are calculated for 2, 5, 10, 15, 20 and 25 \msun (Table \ref{t:binary}) progenitor masses.

\begin{table*}
\caption{The binary separations, the orbital velocities of the progenitor, the orbital periods
and the Roche Lobe radii of the pre--supernova binary for various final masses of the progenitor with 13 \msun HD\,37424 are given.
Roche Lobe radii are calculated based on
\citet{1983ApJ...268..368E}.}
\label{t:binary}
\begin{center}
\begin{tabular}{c r@{~}l r@{~}l r@{~}l r@{~}l r@{~}l r@{~}l}
Progenitor Mass (\msun)
& \multicolumn{2}{c}{ 2}
& \multicolumn{2}{c}{ 5}
& \multicolumn{2}{c}{10}
& \multicolumn{2}{c}{15}
& \multicolumn{2}{c}{20}
& \multicolumn{2}{c}{25} \\\hline
\noalign{\smallskip}
Binary Separation (\rsun)
&    9 & $\pmpm{-1}{+3}$
&   49 & $\pmpm{-9}{+11}$
&  152 & $\pmpm{-26}{+37}$
&  281 & $\pmpm{-49}{+68}$
&  425 & $\pmpm{-75}{+101}$
&  576 & $\pmpm{-101}{+137}$ \\
\noalign{\smallskip}
Orbital Velocity (\kms)
&  481 & $\pm$49
&  192 & $\pm$20
&   96 & $\pm$10
&   64 & $\pm$7
&   48 & $\pm$5
&   38 & $\pm$4 \\
\noalign{\smallskip}
Orbital Period (days)
& 0.85 & $\pmpm{-0.22}{+0.32}$
&    9 & $\pmpm{-2}{+4}$
&   45 & $\pmpm{-11}{+17}$
&  103 & $\pmpm{-26}{+39}$
&  176 & $\pmpm{-44}{+66}$
&  259 & $\pmpm{-65}{+98}$ \\
\noalign{\smallskip}
Roche Lobe Radius (\rsun)
&      &
&      &
&      &
&  110 & $\pmpm{-19}{+27}$
&  177 & $\pmpm{-31}{+42}$
&  251 & $\pmpm{-44}{+60}$ \\\hline
\noalign{\smallskip}
\end{tabular}
\end{center}
\end{table*}

As discussed in the previous section, the OB stars around might be members of an unidentified old OB association
of which all of the O type stars underwent supernova explosions.
This also makes a plausible explanation for the low density medium in which the SNR expands symmetrically.
But, an ejection of the pre-SN system is also possible.
A membership to an OB association or to a cluster is important also regarding the distance determination.
In this work, the distance derived from pulsar parallax ($1.3\pmpm{+0.20}{-0.16}$ \kpc) is accepted as the most reliable estimation.
Also, the distance to the star measured from interstellar lines is in the same range, $1288\pmpm{+304}{-193}$ (see section 3).
The spectro--photometric distance is much larger by using absolute magnitudes from \citet{1982lbg6.conf.....A}.
Yet,
by using typical luminosities for B0.5V type suggested in
\citet{2010AN....331..349H},
it is 1318$\pm$119 \pc.
Hence, the distance to the star and the SNR can be assumed to be 1.3 \kpc.
However, the \Av measured directly towards S147 is much lower than the \Av towards the stars beyond 1 \kpc.
Furthermore, two stars, HD\,36665 and HD\,37318, show highly shifted interstellar \ion{Ca}{ii} and \ion{Na}{i} lines
related to the SNR implying that these objects are background sources
\citep{2004A&A...426..555S}.
Their distances based on the reported spectral types are closer to the Sun than HD\,37424 is.
HD\,36665 has $837\pmpm{+347}{-285}$ \pc for B1V type and HD\,37318 is $903\pmpm{+231}{-228}$ \pc far away adopting B0.5V.
On the other hand, the distance for HD\,36665 is identified as 1860 \pc in
\citet{1980BICDS..19...61N} through H$_{\beta}$ measurements.
It has a high \Av of 1.74 mag.
If we assume that HD\,36665 is also at 1.3$\pm$0.1 \kpc,
then it must be a bright B1V type star which is 1--1.5 mag brighter than the average value given in W06.
HD\,37318 also can be a member of the possible OB association.

The supernova event had occurred quite nearby in a fairly reddened medium.
Assuming a very faint SN (\Mv = $-14$ mag), the apparent visual magnitude is $-2.1$ mag, and
for a bright SN with \Mv$ = -21$ mag, the apparent visual magnitude is $-9.1$ mag; as bright as SN\,1006.
The event might have also caused the $^{10}\mathrm{Be}$ peak at 35$\pm$2 \kyr measured in
\citet{1987Natur.326..273R}
from the deep ice cores from Dome C and Vostok Antarctica.
\section{Conclusion}

HD\,37424 is a B0.5V type runaway star from a binary ejection due to the supernova which gave birth to SNR S147 and PSR\,J0538+2817.
The star has $-9.2\pm6.5$ \kms heliocentric radial velocity and 74$\pm$8 \kms 3--D peculiar velocity.
The distance calculated from interstellar \ion{Ca}{ii} lines is $1288\pmpm{+304}{-193}$ \pc and from spectro--photometry 1318$\pm$119 \pc.
No high velocity gas related with the SNR is detected in the spectra.
The past trajectories of the pulsar and HD\,37424 are reconstructed throughout Monte Carlo simulations.
It is found that both stars were at the same position at
$\left(l, b\right)=\left(84.82\pm0.01, 27.84\pm0.01\right)$ deg at $30\pm4$ \kyr in the past.
The position of the explosion is $4.2\pmpm{+0.8}{-0.6}$\,arcmin away from the geometrical center.
The distance of the SNR is found as 1333$\pmpm{+103}{-112}$ \pc.
\Av towards the SNR is 1.28$\pm$0.06 mag.

Today's kinematics of the stars are well known,
further detailed calculations should be done to find the true kick vectors of the pulsar and a possible spin-kick alignment.

There is no known OB association reported close to the SNR.
19 OB type stars within 100\pc from the SNR geometrical center have very similar 2--D velocities with the 5 runaway stars including HD\,37424.
This might be an old OB association having no bright O type stars.
The low density medium in which the SNR is expanding is probably due to the previous SNe and H II regions
driven by this old association.
The progenitor was a massive star with a zero age main sequence mass greater than 13 \msun.
Considering the lack of O type stars in the field, a progenitor mass much larger than 20 \msun is not expected.
Assuming a circular orbit, the pre--supernova binary separation is in the range of 8 to 711 \rsun for 2 to 25 \msun final mass of the progenitor.
The corresponding Roche Lobe radii for 15 to 25 \msun masses vary from 91 to 311 \rsun.
It must have been an interacting binary.

For 1.3 kpc distance and 1.3 mag extinction, the SN which happened 30$\pm$4 kyr ago had an apparent brightness of $-2.1$ to $-9.1$ mag.

The source will be investigated regarding the elemental signatures of binary accretion and
the possible supernova debris on its photosphere through high resolution and high S/N spectra.
\section*{Acknowledgements}
This work was conceived by the late Oktay H. GUSE\.{I}NOV (1938-2009).
The authors acknowledge the leading role he had in all stages of this work and other related topics.
His contribution to Galactic Astrophysics and his scholarship will be greatly missed.
We thank Janos Schmidt and Christian Ginski for their observational support and Ronny Errmann for useful discussions.
BD, NT and RN acknowledge support from DFG in the SFB/TR-7 Gravitational Wave Astronomy.
This research has made use of Aladin.
We would like to thank the Calar Alto observatory staff
for their help in our observations;
and we would like to thank DFG in NE 515 / 53-1 for
financial support for the Calar Alto observing run.
We also thank \tubitak National Observatory for their supports in using RTT--150 with project number 12ARTT150-267.


  \bsp
  \label{lastpage}

\begin{thebibliography}{}

\bibitem[\protect\citeauthoryear{{Allakhverdiev}, {Guseinov}, {Tagieva} \&
  {Yusifov}}{{Allakhverdiev} et~al.}{1997}]{1997ARep...41..257A}
{Allakhverdiev} A.~O.,  {Guseinov} O.~H.,  {Tagieva} S.~O.,    {Yusifov} I.~M.,
   1997, Astronomy Reports, 41, 257

\bibitem[\protect\citeauthoryear{{Aller} et~al.,}{{Aller}
  et~al.}{1982}]{1982lbg6.conf.....A}
{Aller} L.~H.  et~al., eds, 1982, {Landolt-B{\"o}rnstein: Numerical Data and
  Functional Relationships in Science and Technology - New Series ''
  Gruppe/Group 6 Astronomy and Astrophysics '' Volume 2 Schaifers/Voigt:
  Astronomy and Astrophysics / Astronomie und Astrophysik '' Stars and Star
  Clusters / Sterne und Sternhaufen}

\bibitem[\protect\citeauthoryear{{Anderson}, {Cadwell}, {Jacoby}, {Wolszczan},
  {Foster} \& {Kramer}}{{Anderson} et~al.}{1996}]{1996ApJ...468L..55A}
{Anderson} S.~B.,  {Cadwell} B.~J.,  {Jacoby} B.~A.,  {Wolszczan} A.,  {Foster}
  R.~S.,    {Kramer} M.,  1996, ApJ, 468, L55

\bibitem[\protect\citeauthoryear{{Ankay}, {Kaper}, {de Bruijne}, {Dewi},
  {Hoogerwerf} \& {Savonije}}{{Ankay} et~al.}{2001}]{2001A&A...370..170A}
{Ankay} A.,  {Kaper} L.,  {de Bruijne} J.~H.~J.,  {Dewi} J.,  {Hoogerwerf} R.,
    {Savonije} G.~J.,  2001, A\&A, 370, 170

\bibitem[\protect\citeauthoryear{{Blaauw}}{{Blaauw}}{1961}]{1961BAN....15..265%
B}
{Blaauw} A.,  1961, Bull.~of Astro.\ Inst.\ of the Netherlands, 15, 265

\bibitem[\protect\citeauthoryear{{Bobylev}}{{Bobylev}}{2008}]{2008AstL...34..6%
86B}
{Bobylev} V.~V.,  2008, Astronomy Letters, 34, 686

\bibitem[\protect\citeauthoryear{{Bouigue}, {Boulon} \& {Pedoussaut}}{{Bouigue}
  et~al.}{1961}]{1961AnTou..28...33B}
{Bouigue} R.,  {Boulon} J.,    {Pedoussaut} A.,  1961, Annales de
  l'Observatoire Astron.~et Meteo.~de Toulouse, 28, 33

\bibitem[\protect\citeauthoryear{{Chatterjee} et~al.,}{{Chatterjee}
  et~al.}{2009}]{2009ApJ...698..250C}
{Chatterjee} S.  et~al., 2009, ApJ, 698, 250

\bibitem[\protect\citeauthoryear{{Chevalier}}{{Chevalier}}{1974}]{1974ApJ...18%
8..501C}
{Chevalier} R.~A.,  1974, ApJ, 188, 501

\bibitem[\protect\citeauthoryear{{Christy}}{{Christy}}{1977}]{1977ApJ...217..1%
27C}
{Christy} J.~W.,  1977, ApJ, 217, 127

\bibitem[\protect\citeauthoryear{{Clark} \& {Caswell}}{{Clark} \&
  {Caswell}}{1976}]{1976MNRAS.174..267C}
{Clark} D.~H.,  {Caswell} J.~L.,  1976, MNRAS, 174, 267

\bibitem[\protect\citeauthoryear{{Clausen} \& {Jensen}}{{Clausen} \&
  {Jensen}}{1979}]{1979RA......9..479C}
{Clausen} J.~V.,  {Jensen} K.~S.,  1979, in {McCarthy} M.~F.,  {Philip}
  A.~G.~D.,   {Coyne} G.~V.,  eds,  Ricerche Astronomiche Vol. 9, IAU Colloq.
  47: Spectral Classification of the Future. p.~479

\bibitem[\protect\citeauthoryear{{Cordes} \& {Lazio}}{{Cordes} \&
  {Lazio}}{2002}]{2002astro.ph..7156C}
{Cordes} J.~M.,  {Lazio} T.~J.~W.,  2002, ArXiv Astrophysics e-prints

\bibitem[\protect\citeauthoryear{{Cucchiaro}, {Jaschek}, {Jaschek} \&
  {Macau-Hercot}}{{Cucchiaro} et~al.}{1976}]{1976A+AS...26..241C}
{Cucchiaro} A.,  {Jaschek} M.,  {Jaschek} C.,    {Macau-Hercot} D.,  1976,
  A\&AS, 26, 241

\bibitem[\protect\citeauthoryear{{Eggleton}}{{Eggleton}}{1983}]{1983ApJ...268.%
.368E}
{Eggleton} P.~P.,  1983, ApJ, 268, 368

\bibitem[\protect\citeauthoryear{{Feast} \& {Shuttleworth}}{{Feast} \&
  {Shuttleworth}}{1965}]{1965MNRAS.130..245F}
{Feast} M.~W.,  {Shuttleworth} M.,  1965, MNRAS, 130, 245

\bibitem[\protect\citeauthoryear{{Fesen}, {Blair} \& {Kirshner}}{{Fesen}
  et~al.}{1985}]{1985ApJ...292...29F}
{Fesen} R.~A.,  {Blair} W.~P.,    {Kirshner} R.~P.,  1985, ApJ, 292, 29

\bibitem[\protect\citeauthoryear{{Fuerst} \& {Reich}}{{Fuerst} \&
  {Reich}}{1986}]{1986A&A...163..185F}
{Fuerst} E.,  {Reich} W.,  1986, A\&A, 163, 185

\bibitem[\protect\citeauthoryear{{Gaensler}, {Stappers}, {Frail}, {Moffett},
  {Johnston} \& {Chatterjee}}{{Gaensler} et~al.}{2000}]{2000MNRAS.318...58G}
{Gaensler} B.~M.,  {Stappers} B.~W.,  {Frail} D.~A.,  {Moffett} D.~A.,
  {Johnston} S.,    {Chatterjee} S.,  2000, MNRAS, 318, 58

\bibitem[\protect\citeauthoryear{{Gies}}{{Gies}}{1987}]{1987ApJS...64..545G}
{Gies} D.~R.,  1987, ApJS, 64, 545

\bibitem[\protect\citeauthoryear{{Gies} \& {Bolton}}{{Gies} \&
  {Bolton}}{1986}]{1986ApJS...61..419G}
{Gies} D.~R.,  {Bolton} C.~T.,  1986, ApJS, 61, 419

\bibitem[\protect\citeauthoryear{{Green}}{{Green}}{2009}]{2009yCat.7253....0G}
{Green} D.~A.,  2009, VizieR Online Data Catalog, 7253, 0

\bibitem[\protect\citeauthoryear{{Guseinov}, {Ankay}, {Sezer} \&
  {Tagieva}}{{Guseinov} et~al.}{2003}]{2003A&AT...22..273G}
{Guseinov} O.~H.,  {Ankay} A.,  {Sezer} A.,    {Tagieva} S.~O.,  2003,
  Astronomical and Astrophysical Transactions, 22, 273

\bibitem[\protect\citeauthoryear{{Guseinov}, {Ankay} \& {Tagieva}}{{Guseinov}
  et~al.}{2003}]{2003SerAJ.167...93G}
{Guseinov} O.~H.,  {Ankay} A.,    {Tagieva} S.~O.,  2003, Serbian Astronomical
  Journal, 167, 93

\bibitem[\protect\citeauthoryear{{Guseinov}, {Ankay} \& {Tagieva}}{{Guseinov}
  et~al.}{2004a}]{2004SerAJ.168...55G}
{Guseinov} O.~H.,  {Ankay} A.,    {Tagieva} S.~O.,  2004a, Serbian Astronomical
  Journal, 168, 55

\bibitem[\protect\citeauthoryear{{Guseinov}, {Ankay} \& {Tagieva}}{{Guseinov}
  et~al.}{2004b}]{2004SerAJ.169...65G}
{Guseinov} O.~H.,  {Ankay} A.,    {Tagieva} S.~O.,  2004b, Serbian Astronomical
  Journal, 169, 65

\bibitem[\protect\citeauthoryear{{Guseinov}, {Ankay} \& {Tagieva}}{{Guseinov}
  et~al.}{2004c}]{2004IJMPD..13.1805G}
{Guseinov} O.~H.,  {Ankay} A.,    {Tagieva} S.~O.,  2004c, International
  Journal of Modern Physics D, 13, 1805

\bibitem[\protect\citeauthoryear{{Guseinov}, {Ankay} \& {Tagieva}}{{Guseinov}
  et~al.}{2005}]{2005Ap.....48..330G}
{Guseinov} O.~H.,  {Ankay} A.,    {Tagieva} S.~O.,  2005, Astrophysics, 48, 330

\bibitem[\protect\citeauthoryear{{Hansen} \& {Phinney}}{{Hansen} \&
  {Phinney}}{1997}]{1997MNRAS.291..569H}
{Hansen} B.~M.~S.,  {Phinney} E.~S.,  1997, MNRAS, 291, 569

\bibitem[\protect\citeauthoryear{{Herbig}}{{Herbig}}{1975}]{1975ApJ...196..129%
H}
{Herbig} G.~H.,  1975, ApJ, 196, 129

\bibitem[\protect\citeauthoryear{{Hobbs}, {Lorimer}, {Lyne} \&
  {Kramer}}{{Hobbs} et~al.}{2005}]{2005MNRAS.360..974H}
{Hobbs} G.,  {Lorimer} D.~R.,  {Lyne} A.~G.,    {Kramer} M.,  2005, MNRAS, 360,
  974

\bibitem[\protect\citeauthoryear{{Hohle}, {Neuh{\"a}user} \& {Schutz}}{{Hohle}
  et~al.}{2010}]{2010AN....331..349H}
{Hohle} M.~M.,  {Neuh{\"a}user} R.,    {Schutz} B.~F.,  2010, Astronomische
  Nachrichten, 331, 349

\bibitem[\protect\citeauthoryear{{Hoogerwerf}, {de Bruijne} \& {de
  Zeeuw}}{{Hoogerwerf} et~al.}{2001}]{2001A&A...365...49H}
{Hoogerwerf} R.,  {de Bruijne} J.~H.~J.,    {de Zeeuw} P.~T.,  2001, A\&A, 365,
  49

\bibitem[\protect\citeauthoryear{{Hunter} \& {Massey}}{{Hunter} \&
  {Massey}}{1990}]{1990AJ.....99..846H}
{Hunter} D.~A.,  {Massey} P.,  1990, AJ, 99, 846

\bibitem[\protect\citeauthoryear{{Kaper}, {van Loon}, {Augusteijn},
  {Goudfrooij}, {Patat}, {Waters} \& {Zijlstra}}{{Kaper}
  et~al.}{1997}]{1997ApJ...475L..37K}
{Kaper} L.,  {van Loon} J.~T.,  {Augusteijn} T.,  {Goudfrooij} P.,  {Patat} F.,
   {Waters} L.~B.~F.~M.,    {Zijlstra} A.~A.,  1997, ApJ, 475, L37

\bibitem[\protect\citeauthoryear{{Katsuta} et~al.,}{{Katsuta}
  et~al.}{2012}]{2012ApJ...752..135K}
{Katsuta} J.  et~al., 2012, ApJ, 752, 135

\bibitem[\protect\citeauthoryear{{Kenyon} \& {Hartmann}}{{Kenyon} \&
  {Hartmann}}{1995}]{1995ApJS..101..117K}
{Kenyon} S.~J.,  {Hartmann} L.,  1995, ApJS, 101, 117

\bibitem[\protect\citeauthoryear{{Kharchenko}}{{Kharchenko}}{2001}]{2001KFNT..%
.17..409K}
{Kharchenko} N.~V.,  2001, Kinematika i Fizika Nebesnykh Tel, 17, 409

\bibitem[\protect\citeauthoryear{{Kirshner} \& {Arnold}}{{Kirshner} \&
  {Arnold}}{1979}]{1979ApJ...229..147K}
{Kirshner} R.~P.,  {Arnold} C.~N.,  1979, ApJ, 229, 147

\bibitem[\protect\citeauthoryear{{Kramer}, {Lyne}, {Hobbs}, {L{\"o}hmer},
  {Carr}, {Jordan} \& {Wolszczan}}{{Kramer} et~al.}{2003}]{2003ApJ...593L..31K}
{Kramer} M.,  {Lyne} A.~G.,  {Hobbs} G.,  {L{\"o}hmer} O.,  {Carr} P.,
  {Jordan} C.,    {Wolszczan} A.,  2003, ApJ, 593, L31

\bibitem[\protect\citeauthoryear{{Kundu}, {Angerhofer}, {Fuerst} \&
  {Hirth}}{{Kundu} et~al.}{1980}]{1980A&A....92..225K}
{Kundu} M.~R.,  {Angerhofer} P.~E.,  {Fuerst} E.,    {Hirth} W.,  1980, A\&A,
  92, 225

\bibitem[\protect\citeauthoryear{{Lozinskaya}}{{Lozinskaya}}{1976}]{1976SvA...%
.20...19L}
{Lozinskaya} T.~A.,  1976, SvA, 20, 19

\bibitem[\protect\citeauthoryear{{Lyne} \& {Lorimer}}{{Lyne} \&
  {Lorimer}}{1994}]{1994Natur.369..127L}
{Lyne} A.~G.,  {Lorimer} D.~R.,  1994, Nat, 369, 127

\bibitem[\protect\citeauthoryear{{McGowan}, {Kennea}, {Zane}, {C{\'o}rdova},
  {Cropper}, {Ho}, {Sasseen} \& {Vestrand}}{{McGowan}
  et~al.}{2003}]{2003ApJ...591..380M}
{McGowan} K.~E.,  {Kennea} J.~A.,  {Zane} S.,  {C{\'o}rdova} F.~A.,  {Cropper}
  M.,  {Ho} C.,  {Sasseen} T.,    {Vestrand} W.~T.,  2003, ApJ, 591, 380

\bibitem[\protect\citeauthoryear{{Megier}, {Strobel}, {Galazutdinov} \&
  {Kre{\l}owski}}{{Megier} et~al.}{2009}]{2009A&A...507..833M}
{Megier} A.,  {Strobel} A.,  {Galazutdinov} G.~A.,    {Kre{\l}owski} J.,  2009,
  A\&A, 507, 833

\bibitem[\protect\citeauthoryear{{Milne}}{{Milne}}{1979}]{1979AuJPh..32...83M}
{Milne} D.~K.,  1979, Australian Journal of Physics, 32, 83

\bibitem[\protect\citeauthoryear{{Minkowski}}{{Minkowski}}{1958}]{1958RvMP...3%
0.1048M}
{Minkowski} R.,  1958, Reviews of Modern Physics, 30, 1048

\bibitem[\protect\citeauthoryear{{Morgan}, {Code} \& {Whitford}}{{Morgan}
  et~al.}{1955}]{1955ApJS....2...41M}
{Morgan} W.~W.,  {Code} A.~D.,    {Whitford} A.~E.,  1955, ApJS, 2, 41

\bibitem[\protect\citeauthoryear{{Mugrauer}, {Avila} \& {Guirao}}{{Mugrauer}
  et~al.}{2014}]{2014AN....335..417M}
{Mugrauer} M.,  {Avila} G.,    {Guirao} C.,  2014, Astronomische Nachrichten,
  335, 417

\bibitem[\protect\citeauthoryear{{Mugrauer} \& {Berthold}}{{Mugrauer} \&
  {Berthold}}{2010}]{2010AN....331..449M}
{Mugrauer} M.,  {Berthold} T.,  2010, Astronomische Nachrichten, 331, 449

\bibitem[\protect\citeauthoryear{{Neckel}, {Klare} \& {Sarcander}}{{Neckel}
  et~al.}{1980}]{1980BICDS..19...61N}
{Neckel} T.,  {Klare} G.,    {Sarcander} M.,  1980, Bulletin d'Information du
  Centre de Donnees Stellaires, 19, 61

\bibitem[\protect\citeauthoryear{{Ng}, {Romani}, {Brisken}, {Chatterjee} \&
  {Kramer}}{{Ng} et~al.}{2007}]{2007ApJ...654..487N}
{Ng} C.-Y.,  {Romani} R.~W.,  {Brisken} W.~F.,  {Chatterjee} S.,    {Kramer}
  M.,  2007, ApJ, 654, 487

\bibitem[\protect\citeauthoryear{{Philp}, {Evans}, {Leonard} \&
  {Frail}}{{Philp} et~al.}{1996}]{1996AJ....111.1220P}
{Philp} C.~J.,  {Evans} C.~R.,  {Leonard} P.~J.~T.,    {Frail} D.~A.,  1996,
  AJ, 111, 1220

\bibitem[\protect\citeauthoryear{{Poveda}, {Ruiz} \& {Allen}}{{Poveda}
  et~al.}{1967}]{1967BOTT....4...86P}
{Poveda} A.,  {Ruiz} J.,    {Allen} C.,  1967, Boletin de los Observatorios
  Tonantzintla y Tacubaya, 4, 86

\bibitem[\protect\citeauthoryear{{Przybilla}, {Nieva}, {Heber} \&
  {Butler}}{{Przybilla} et~al.}{2008}]{2008ApJ...684L.103P}
{Przybilla} N.,  {Nieva} M.~F.,  {Heber} U.,    {Butler} K.,  2008, ApJ, 684,
  L103

\bibitem[\protect\citeauthoryear{{Raisbeck}, {Yiou}, {Bourles}, {Lorius} \&
  {Jouzel}}{{Raisbeck} et~al.}{1987}]{1987Natur.326..273R}
{Raisbeck} G.~M.,  {Yiou} F.,  {Bourles} D.,  {Lorius} C.,    {Jouzel} J.,
  1987, Nat, 326, 273

\bibitem[\protect\citeauthoryear{{Rieke} \& {Lebofsky}}{{Rieke} \&
  {Lebofsky}}{1985}]{1985ApJ...288..618R}
{Rieke} G.~H.,  {Lebofsky} M.~J.,  1985, ApJ, 288, 618

\bibitem[\protect\citeauthoryear{{Sallmen} \& {Welsh}}{{Sallmen} \&
  {Welsh}}{2004}]{2004A&A...426..555S}
{Sallmen} S.,  {Welsh} B.~Y.,  2004, A\&A, 426, 555

\bibitem[\protect\citeauthoryear{{Sauvageot}, {Ballet} \&
  {Rothenflug}}{{Sauvageot} et~al.}{1990}]{1990A&A...227..183S}
{Sauvageot} J.~L.,  {Ballet} J.,    {Rothenflug} R.,  1990, A\&A, 227, 183

\bibitem[\protect\citeauthoryear{{Sayer}, {Nice} \& {Kaspi}}{{Sayer}
  et~al.}{1996}]{1996ApJ...461..357S}
{Sayer} R.~W.,  {Nice} D.~J.,    {Kaspi} V.~M.,  1996, ApJ, 461, 357

\bibitem[\protect\citeauthoryear{{Silk} \& {Wallerstein}}{{Silk} \&
  {Wallerstein}}{1973}]{1973ApJ...181..799S}
{Silk} J.,  {Wallerstein} G.,  1973, ApJ, 181, 799

\bibitem[\protect\citeauthoryear{{Skiff}}{{Skiff}}{2013}]{2013yCat....1.2023S}
{Skiff} B.~A.,  2013, VizieR Online Data Catalog, 1, 2023

\bibitem[\protect\citeauthoryear{{Sofue}, {Furst} \& {Hirth}}{{Sofue}
  et~al.}{1980}]{1980PASJ...32....1S}
{Sofue} Y.,  {Furst} E.,    {Hirth} W.,  1980, PASJ, 32, 1

\bibitem[\protect\citeauthoryear{{Steele}, {Negueruela} \& {Clark}}{{Steele}
  et~al.}{1999}]{1999A+AS..137..147S}
{Steele} I.~A.,  {Negueruela} I.,    {Clark} J.~S.,  1999, A\&AS, 137, 147

\bibitem[\protect\citeauthoryear{{Tauris} \& {Takens}}{{Tauris} \&
  {Takens}}{1998}]{1998A&A...330.1047T}
{Tauris} T.~M.,  {Takens} R.~J.,  1998, A\&A, 330, 1047

\bibitem[\protect\citeauthoryear{{Tetzlaff}, {Din{\c c}el}, {Neuh{\"a}user} \&
  {Kovtyukh}}{{Tetzlaff} et~al.}{2014}]{2014MNRAS.438.3587T}
{Tetzlaff} N.,  {Din{\c c}el} B.,  {Neuh{\"a}user} R.,    {Kovtyukh} V.~V.,
  2014, MNRAS, 438, 3587

\bibitem[\protect\citeauthoryear{{Tetzlaff}, {Eisenbeiss}, {Neuh{\"a}user} \&
  {Hohle}}{{Tetzlaff} et~al.}{2011}]{2011MNRAS.417..617T}
{Tetzlaff} N.,  {Eisenbeiss} T.,  {Neuh{\"a}user} R.,    {Hohle} M.~M.,  2011,
  MNRAS, 417, 617

\bibitem[\protect\citeauthoryear{{Tetzlaff}, {Neuh{\"a}user} \&
  {Hohle}}{{Tetzlaff} et~al.}{2011}]{2011MNRAS.410..190T}
{Tetzlaff} N.,  {Neuh{\"a}user} R.,    {Hohle} M.~M.,  2011, MNRAS, 410, 190

\bibitem[\protect\citeauthoryear{{Tetzlaff}, {Neuh{\"a}user}, {Hohle} \&
  {Maciejewski}}{{Tetzlaff} et~al.}{2010}]{2010MNRAS.402.2369T}
{Tetzlaff} N.,  {Neuh{\"a}user} R.,  {Hohle} M.~M.,    {Maciejewski} G.,  2010,
  MNRAS, 402, 2369

\bibitem[\protect\citeauthoryear{{Tetzlaff}, {Schmidt}, {Hohle} \&
  {Neuh{\"a}user}}{{Tetzlaff} et~al.}{2012}]{2012PASA...29...98T}
{Tetzlaff} N.,  {Schmidt} J.~G.,  {Hohle} M.~M.,    {Neuh{\"a}user} R.,  2012,
  29, 98

\bibitem[\protect\citeauthoryear{{Tetzlaff}, {Torres}, {Neuh{\"a}user} \&
  {Hohle}}{{Tetzlaff} et~al.}{2013}]{2013MNRAS.435..879T}
{Tetzlaff} N.,  {Torres} G.,  {Neuh{\"a}user} R.,    {Hohle} M.~M.,  2013,
  MNRAS, 435, 879

\bibitem[\protect\citeauthoryear{{van den Bergh}}{{van den
  Bergh}}{1980}]{1980JApA....1...67V}
{van den Bergh} S.,  1980, Journal of Astrophysics and Astronomy, 1, 67

\bibitem[\protect\citeauthoryear{{van den Heuvel}}{{van den
  Heuvel}}{1993}]{1993SSRv...66..309V}
{van den Heuvel} E.~P.~J.,  1993, 66, 309

\bibitem[\protect\citeauthoryear{{Walborn}}{{Walborn}}{1971}]{1971ApJS...23..2%
57W}
{Walborn} N.~R.,  1971, ApJS, 23, 257

\bibitem[\protect\citeauthoryear{{Walborn} \& {Fitzpatrick}}{{Walborn} \&
  {Fitzpatrick}}{1990}]{1990PASP..102..379W}
{Walborn} N.~R.,  {Fitzpatrick} E.~L.,  1990, PASP, 102, 379

\bibitem[\protect\citeauthoryear{{Wang} \& {Gies}}{{Wang} \&
  {Gies}}{1998}]{1998PASP..110.1310W}
{Wang} Z.~X.,  {Gies} D.~R.,  1998, PASP, 110, 1310

\bibitem[\protect\citeauthoryear{{Wegner}}{{Wegner}}{1994}]{1994MNRAS.270..229%
W}
{Wegner} W.,  1994, MNRAS, 270, 229

\bibitem[\protect\citeauthoryear{{Wegner}}{{Wegner}}{2006}]{2006MNRAS.371..185%
W}
{Wegner} W.,  2006, MNRAS, 371, 185

\bibitem[\protect\citeauthoryear{{Wongwathanarat}, {Janka} \&
  {M{\"u}ller}}{{Wongwathanarat} et~al.}{2013}]{2013A&A...552A.126W}
{Wongwathanarat} A.,  {Janka} H.-T.,    {M{\"u}ller} E.,  2013, A\&A, 552, A126

\bibitem[\protect\citeauthoryear{{Woosley}, {Langer} \& {Weaver}}{{Woosley}
  et~al.}{1995}]{1995ApJ...448..315W}
{Woosley} S.~E.,  {Langer} N.,    {Weaver} T.~A.,  1995, ApJ, 448, 315

\bibitem[\protect\citeauthoryear{{Zacharias}, {Finch}, {Girard}, {Henden},
  {Bartlett}, {Monet} \& {Zacharias}}{{Zacharias}
  et~al.}{2012}]{2012yCat.1322....0Z}
{Zacharias} N.,  {Finch} C.~T.,  {Girard} T.~M.,  {Henden} A.,  {Bartlett}
  J.~L.,  {Monet} D.~G.,    {Zacharias} M.~I.,  2012, VizieR Online Data
  Catalog, 1322, 0

\bibitem[\protect\citeauthoryear{{Zavlin} \& {Pavlov}}{{Zavlin} \&
  {Pavlov}}{2004}]{2004MmSAI..75..458Z}
{Zavlin} V.~E.,  {Pavlov} G.~G.,  2004, 75, 458

\end{thebibliography}
\end{document}